\documentclass[a4paper,11pt]{article}
\usepackage{jcappub} 

\usepackage[export]{adjustbox} 
\usepackage[utf8]{inputenc}
\usepackage{color}
\usepackage{graphicx}
\usepackage{amsmath,amsfonts,amssymb}
\usepackage{xcolor}
\usepackage{tikz}
\usepackage{booktabs}
\usepackage{url}
\usepackage{bm}
\usepackage{caption}
\definecolor{gpcolor}{RGB}{0,0,255}

\title{From the Solar System to cosmological distances: a complete formalism for gravitational wave astrometry}

\author[a]{Gabriele Perna,}
\author[b,c,d]{Nicola Bellomo,}
\author[b]{Vincenzo Roatti,}
\author[b,c,d]{Daniele Bertacca}

\affiliation[a]{Keemilise ja Bioloogilise F\"u\"usika Instituut, R\"avala pst. 10, 10143 Tallinn, Estonia}
\affiliation[b]{Dipartimento di Fisica e Astronomia ``Galileo Galilei'', Universit\`a degli Studi di Padova, Via Marzolo 8, I-35131, Padova, Italy.}
\affiliation[c]{INFN, Sezione di Padova, Via Marzolo 8, I-35131, Padova, Italy.}
\affiliation[d]{INAF, Osservatorio astronomico di Padova, Vicolo dell'Osservatorio 5, I-35122 Padova, Italy.}

\emailAdd{gabriele.perna@phd.unipd.it}
\emailAdd{nicola.bellomo@unipd.it}
\emailAdd{vincenzo.roatti@phd.unipd.it}
\emailAdd{daniele.bertacca@unipd.it}

\abstract{
The presence of a gravitational wave background (GWB) can be established not only via exquisitely precise pulsar timing array (PTA) measurements, but also via astrometric observations. Indeed, the very same background responsible for the delay in the arrival time of pulse causes an apparent displacement of galactic objects as stars and asteroids.
In this work we provide a framework that allows to derive the displacement of sources overcoming the usually adopted ``infinite distance'' approximation.
We also present how this formalism can be used to study the displacements of objects at distances comparable to the GW wavelength, as asteroids, and of objects with a non-trivial three-dimensional distribution, as stars in the Milky Way. Thus, it can be used to probe frequencies beyond PTA experiments, reaching the mHz GWs, also detectable by LISA. We forecast the capability of observing the astrometric deflection induced by a GWB evaluating the harmonic signal-to-noise ratio including correlations between different probes. We find an SNR greater than one for the relevant cases considered and as a consequence a promising Fisher forecast, suggesting a constraining power up to the percent level for a flat background. 
}

\begin{document}
\maketitle

\section{Introduction}
The detection of the first Gravitational Wave (GW) event in 2015 opened a new window to the Universe, providing a further confirmation of General Relativity and insights on mergers of astrophysical binaries~\cite{LIGOScientific:2016vbw, LIGOScientific:2016lio}. 
Currently, one of the main goals of current GW searches is the detection of a stochastic GW background (GWB), whether of cosmological~\cite{Caprini:2018mtu, Guzzetti:2016mkm} or astrophysical origin~\cite{Sathyaprakash:2011bh, LIGOScientific:2016wof, Maggiore:2019uih}. 
The latter would provide further information on the population of astrophysical GW sources~\cite{Ferrari:1998jf, Ferrari:1998ut, Ignatiev:2001jr, 2004MNRAS.351.1237H, Regimbau:2011rp}; whereas the former would shed light on early-Universe processes connected to inflationary dynamics, cosmic strings, or first-order phase transitions~\cite{Bartolo:2016ami, Bartolo:2019yeu, Guzzetti:2016mkm, Caprini:2018mtu, Maggiore:2007ulw}, just to mention few of the most relevant examples. 
In the last couple of years, Pulsar Timing Array (PTA) experiments have already started contributing to this endeavor by finding for the first time some evidence of a Hellings-Downs~\cite{Hellings:1983fr} pattern in the angular correlations of pulsars~\cite{NANOGrav:2023gor, EPTA:2023fyk, Reardon:2023gzh, Xu:2023wog, Miles_2024}, thus, indicating the presence of a GWB.
Unfortunately, the origin of this background is still debated~\cite{NANOGrav:2023hvm}; nevertheless, forthcoming data coming from next generation experiments have the potential to address this question. 
In fact, the characterization of the GWB is one of science goals of the SKAO~\cite{Janssen:2014dka, Keane:2014vja}, as well as of third generation interferometers as LISA~\cite{LISA:2017pwj} in the milli-Hz band, and ET~\cite{Maggiore:2019uih,ET:2025xjr} or CE~\cite{Reitze:2019iox} in the kilo-Hz band. 

One of the primary limitations of current GW searches is the presence of observational gaps in the accessible frequency bands, which motivated the proposal of future detectors such as AEDGE~\cite{AEDGE:2019nxb} and TIANQin~\cite{TianQin:2015yph} to cover the gap between milli- and the kilo-Hz bands, or~$\mu$-ARES~\cite{Sesana:2019vho} and FRB timing~\cite{Lu:2024yuo} to cover the $\mu$-Hz band. 
A separate but promising avenue, which would allow the detection of a GWB across a large portion of the frequency spectrum, is given by astrometric surveys. 
The core idea of this approach is based on the periodic observation of the position of galactic objects, such as stars or asteroids. 
As suggested in refs.~\cite{Linder:1986fdo, Braginsky:1989pv, Damour:1998jm, Pyne:1995iy, Kaiser:1996wk, Jaffe:2004it, Book:2010pf}, GWs perturb the propagation of photons, inducing not only delay in their time of arrival to the detectors, but also a deviation in their trajectory, which is perceived by an observer as a deflection in the angular position of the source.
The frequency range probed by this technique depends on the observation time and the cadence of the survey: large observation times correspond to small minimum frequencies, even reaching the nano-Hz range for measurements that last years; low cadences correspond to large maximum frequency, up to the milli-Hz band.
As an example, Gaia astrometric measurements would be sensitive to GWs in the range~$10^{-8.5}<f<10^{-7.5}$ Hz \cite{Klioner:2017asb, Moore:2017ity}, being able to potentially outperform PTA results, while Nancy Grace Roman could reach up to~$f\sim 10^{-4}$ Hz~\cite{Wang:2020pmf, Wang:2022sxn, Pardo:2023cag}, overlapping with LISA. 
The frequency overlap further highlights the importance of cross-correlating the astrometric and interferometer datasets, enhancing the precision of the GWB measurements and providing a complementary cross-validation of the results.

The literature on this topic has been quite prolific.
The first characterization of the expected astrometric deflection signal due to a GW background was provided by~\cite{Book:2010pf} in the distant-source limit, i.e., assuming the distance to the sources is larger than the observed GW wavelength, and considering the positions of distant quasars~\cite{Pyne:1995iy}. 
This allowed to place the first upper bounds on the dimensionless GW energy density~$\Omega_{\mathrm{GW}}$~\cite{Gwinn:1996gv, Titov_2011, Darling:2018hmc} using this method. 
Similar analyses by~\cite{Aoyama:2021xhj, Wang:2022sxn, Jaraba:2023djs, Caliskan:2023cqm, Zhang:2025srs} have been recently performed using quasars. 
In particular, refs.~\cite{Caliskan:2023cqm,Vaglio:2025tex} provided the first Fisher forecasts on the potential of astrometric surveys, while ref.~\cite{Zhang:2025srs} reports the first Bayesian search for the presence of a coherent GW signal with astrometric measurements. Furthermore, ref.~\cite{Caldarola:2024uig} highlighted the capability of neural networks to infer the GWB from astrometric data.
Finally, in the work by~\cite{Mentasti:2023gmr}, the authors analyzed the short-distance limit in the context of Solar System sources. 

In this work, we overcome the approximations introduced in previous works in the derivation of the angular deflection due to the presence of a GW background, providing for the first time a compact expression valid for any frequency and sources at any distance. 
We then show how the full expression is necessary for objects whose distance is comparable to the GW wavelength, in order not to misestimate the final displacement.
We also produce some astrometry-GW joint analysis forecasts for different GWB models to showcase the improvement that the combination of these two complementary measures provide.

This work is organized as follows.
Section~\ref{sec:probing_gwb} describes the general properties of GWB and the observables analyzed in this work, while in section~\ref{sec:observational_targets} we discuss the observational targets, i.e., astrophysical objects, which are best suited to target such observables.
Section~\ref{sec:forecasts} presents the statistical analysis, while in section~\ref{sec:results} we report our results for different GWB models.
Finally, we conclude in section~\ref{sec:conclusions}.
Appendices~\ref{app:notation_conventions}, \ref{app:explicit_Cl_calculation}, and~\ref{app:population_properties} contain the most theoretical part of the material of the paper, along with the lengthy derivations of our main theoretical results and additional discussion on the properties of the observational targets.
In this work, we use natural units~$c=G=1$.
The chosen metric signature is~$(-,+,+,+)$.
Indices labeled with Greek (Latin) letters run over~$0,1,2,3$ ($1,2,3$).


\section{Probing the gravitational wave background}
\label{sec:probing_gwb}


\subsection{Characterization of the background}

For the purpose of this work, we rely on a description of the GWB that is common to both astrophysical and cosmological backgrounds.
We consider the metric tensor fluctuation~$h_{ij}$ that describes the GWB made up of the linear superposition of individual GW modes as
\begin{equation}
\label{eq::hij}
    h_{ij}(t,\mathbf{x}) = \sum_\lambda \int_{-\infty}^{+\infty} df \int d\hat{\mathbf{p}}\ h_\lambda(f,\hat{\mathbf{p}}) e^\lambda_{ij}(\hat{\mathbf{p}}) e^{-2\pi i f(t-\hat{\mathbf{p}} \cdot \mathbf{x})},
\end{equation}
where~$(t,\mathbf{x})$ indicate the cosmic time and comoving position, $(f,\hat{\mathbf{p}})$ label the GW frequency and direction of propagation, $\lambda$ indicates the polarization degree of freedom, $h_\lambda$ is the mode amplitude for each polarization, and~$e^\lambda_{ij}$ is the polarization tensor.
Fourier transform conventions are explicitly reported in appendix~\ref{app:notation_conventions}.
The variance of metric perturbations in the case of a stationary, unpolarized, and isotropic GWB reads as (see, e.g., ref.~\cite{Caprini:2018mtu})
\begin{equation}
    \left\langle h_\lambda(f,\hat{\mathbf{p}}) h^*_{\lambda'}(f',\hat{\mathbf{p}}') \right\rangle =  \delta^K_{\lambda\lambda'} \delta^D(f-f') \delta^D(\hat{\mathbf{p}}-\hat{\mathbf{p}}') \frac{S_h(f)}{8\pi},
\end{equation}
where the asterisk indicates the complex conjugate, and~$\delta^K$ and~$\delta^D$ are the Kronecker and Dirac delta functions, respectively.
The one-sided GWB power spectrum~$S_h$ is often rewritten as
\begin{equation}
    S_h(f) = \frac{h_c^2(f)}{f} = \frac{3H^2_0}{2 \pi^2 f^3} \Omega_\mathrm{GW}(f),
\end{equation}
where~$h_c(f)$ is the characteristic strain, $H_0$ is the present-day Hubble expansion rate, and ~$\Omega_\mathrm{GWB}(f) = \rho^{-1}_{0c} d\rho_\mathrm{GWB}/d\log f$ is the relative energy density per log frequency, with~$\rho_\mathrm{GWB}$ and~$\rho _{0c}$ being the GWB energy density and today critical density, respectively.


\subsection{Astrometric measurements}

The presence of a GWB perturbs the propagation path of photons, thus it induces an apparent deflection in the observed position of their emitting sources from their original ``true'' position~\cite{Book:2010pf}. 
Therefore, by probing any astrophysical object (asteroids, stars, galaxies, quasars, etc.) position over the course of a long period of time, we can infer their apparent motion induced by GWs.
Suppose a generic source at position~$\mathbf{r}=r\hat{\mathbf{n}}$, where~$r$ is the distance and~$\hat{\mathbf{n}}$ is the line-of-sight, emits a photon with unperturbed frequency~$\bar{\omega}$ that travels to the observer position~$(t,\mathbf{0})$.
The corresponding photon unperturbed world-line reads as
\begin{equation}
    x^\mu (\Lambda) = (t+\bar{\omega} \Lambda, -\bar{\omega} \Lambda \hat{\mathbf{n}}),
\end{equation}
where the affine parameter~$\Lambda$ takes values in the range~$[\Lambda_r,\Lambda_0] = [-r/\bar{\omega},0]$.
However, because of the presence of even a single GW, the photon is observed in the direction~$\hat{\mathbf{n}}+\delta\mathbf{n}$, where the components of the angular displacement are given by~\cite{Book:2010pf}
\begin{equation}
    \delta n^i = \left(\delta^K_{ik} - n^i n^k\right) n^j \left[\frac{1}{2} h_{jk}(\Lambda_0) - \left( \delta^K_{kl} - \frac{p^k n^l}{2(1+\hat{\mathbf{p}}\cdot\hat{\mathbf{n}})} \right) \left( h_{jl}(\Lambda_0) - \frac{1}{\Lambda_r} \int_0^{\Lambda_r} \!\!\!\!\! d\Lambda\ h_{jl}(\Lambda) \right)\right].
\end{equation}
This displacement is orthogonal to the line-of-sight, i.e.,~$\hat{\mathbf{n}} \cdot \delta\mathbf{n}=\mathbf{0}$, as one can immediately check. 
Via a direct substitution of equation~\eqref{eq::hij}, we have that the apparent angular displacement in the position of a single object induced by a GWB reads as
\begin{equation}
    \begin{aligned}
        \delta n^i (t,\mathbf{r}) &= \sum_\lambda \int df d\hat{\mathbf{p}}\ h_\lambda(f,\hat{\mathbf{p}})  e^{-2\pi i f t} \left[ \frac{p^i + n^i}{2(1+\hat{\mathbf{p}}\cdot\hat{\mathbf{n}})} n^j n^k e^\lambda_{jk}(\hat{\mathbf{p}}) - \frac{1}{2} n^j e^\lambda_{ij}(\hat{\mathbf{p}}) \right. \\
        &\qquad\quad \left. + i\mathcal{D} (f,\mathbf{r}) \left( \frac{p^i n^j n^k e^\lambda_{jk}(\hat{\mathbf{p}})}{2(1+\hat{\mathbf{p}}\cdot\hat{\mathbf{n}})} - n^j e^\lambda_{ij}(\hat{\mathbf{p}}) + \frac{(2+\hat{\mathbf{p}}\cdot\hat{\mathbf{n}})}{2(1+\hat{\mathbf{p}}\cdot\hat{\mathbf{n}})} n^i n^j n^k e^\lambda_{jk}(\hat{\mathbf{p}}) \right) \right],
    \end{aligned}
\label{eq:observed_angular_displacement}
\end{equation}
where we introduce the ``distance'' form factor
\begin{equation}
    \mathcal{D} (f,\mathbf{r}) = \frac{e^{2\pi i f r (1+\hat{\mathbf{p}}\cdot\hat{\mathbf{n}})}-1}{2\pi f r(1+\hat{\mathbf{p}}\cdot\hat{\mathbf{n}})}.
\end{equation}
For sources in the ``long-arm limit'', where~$\tilde{x}\equiv2\pi fr \gg 1$, the form factor vanishes and we recover the known ``infinite distance'' result given by the terms in the first line of equation~\eqref{eq:observed_angular_displacement}.
However, the magnitude of the distance form factor for sources not in that regime is substantially different, and induces a non-negligible deviation in the apparent angular displacement of astrophysical objects. 

\begin{figure}[ht] 
\centering
\begin{minipage}[c]{0.6\textwidth}
    \includegraphics[width=\linewidth]{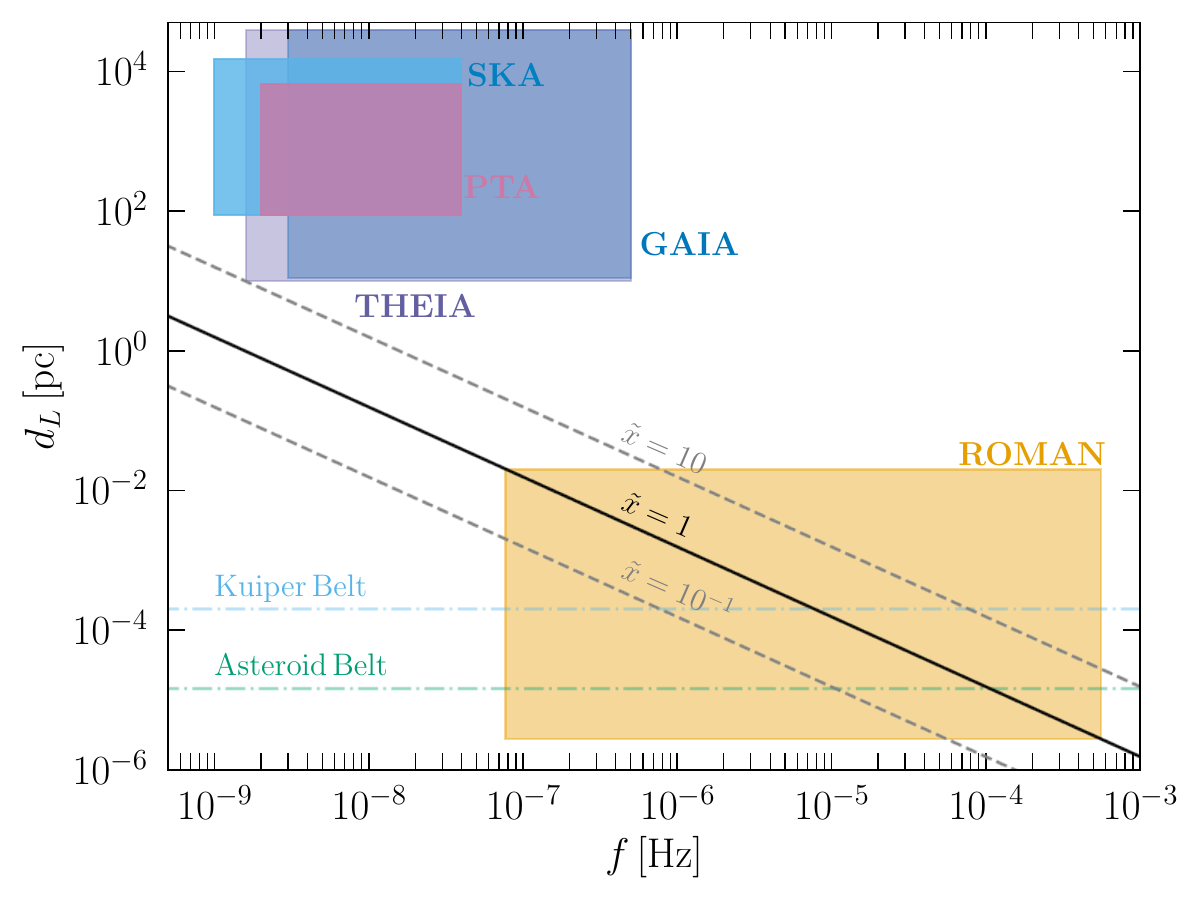}
\end{minipage}
\begin{minipage}[c]{0.8\textwidth}
\resizebox{\textwidth}{!}{ 
\begin{tikzpicture}
\tikzset{every node/.style={font=\Large}} 
\definecolor{bandColor}{RGB}{0, 119, 187}      
\definecolor{gaiaColor}{RGB}{0, 119, 187}      
\definecolor{blueColor}{RGB}{0, 119, 187}      

\definecolor{lightgreen}{RGB}{230, 159, 0}     
\definecolor{darkgreen}{RGB}{230, 159, 0}      

\definecolor{pinkish}{RGB}{204, 121, 167}      
\definecolor{darkred}{RGB}{204, 121, 167}      

\definecolor{orange}{RGB}{213, 94, 0}          
\definecolor{darkorange}{RGB}{86, 180, 233}    

\definecolor{purple}{RGB}{117, 112, 179}       

\definecolor{brown}{RGB}{0, 158, 115}          
\definecolor{cyan}{RGB}{86, 180, 233}          

\definecolor{magenta}{RGB}{94, 60, 153}        

\node[above] at (8,9) {\(\text{Increasing}\quad\Large\tilde{x} \quad \rightarrow\)};

    \draw[thick] (0,7) -- (16,7);
    \draw[thick] (0,4) -- (16,4);
    \draw[thick] (0,1) -- (16,1);
    
    \filldraw[lightgreen, opacity=0.5] (0.2,6.5) rectangle (5,7);
    \filldraw[lightgreen, opacity=0.5] (0.2,3.5) rectangle (5,4);
    \filldraw[lightgreen, opacity=0.5] (0.2,0.5) rectangle (5,1);
    
    \node[text=darkgreen] at (1,6.75) {\textbf{\small {ROMAN}}};
    \node[text=darkgreen] at (1,3.75) {\textbf{\small {ROMAN}}};
    \node[text=darkgreen] at (1,0.75) {\textbf{\small {ROMAN}}};

    \filldraw[cyan, opacity=0.5] (1.7,-0.5) rectangle (5.2,0);
    \filldraw[purple, opacity=0.35] (2,-0.48) rectangle (5.2,-0.02);
    \node[text=purple] at (3.7,-0.25) {\textbf{THEIA}};
    
    \node[text=darkorange] at (2.3,-0.25) {\textbf{\large SKA}};
    \node[text=darkorange] at (2,-0.75) {\textcolor{darkorange}{\(\sim 14 \) days }};
    \node[text=darkorange] at (5.2,-0.75) {\textcolor{darkorange}{\(20\) yr}};   
    
    \filldraw[bandColor, opacity=0.5] (2,6.5) rectangle (5,7);
    \filldraw[bandColor, opacity=0.5] (2,3.5) rectangle (5,4);
    \filldraw[bandColor, opacity=0.5] (2,0.5) rectangle (5,1);

    \node[text=gaiaColor] at (3.3,6.75) {\textbf{\large GAIA}};
    \node[text=gaiaColor] at (3.3,3.75) {\textbf{\large GAIA}};
    \node[text=gaiaColor] at (3.3,0.75) {\textbf{\large GAIA}};

    \filldraw[pinkish, opacity=0.5] (4,6.5) rectangle (5.1,7);
    \filldraw[pinkish, opacity=0.5] (4,3.5) rectangle (5.1,4);
    \filldraw[pinkish, opacity=0.5] (4,0.5) rectangle (5.1,1);

    \node[text=darkred] at (4.5,6.75) {\textbf{\small {PTA}}};
    \node[text=darkred] at (4.5,3.75) {\textbf{\small {PTA}}};
    \node[text=darkred] at (4.5,0.75) {\textbf{\small {PTA}}};

    \node[text=orange] at (5.6, 8.9) {\textbf{Proxima Centauri}};
    \draw[thick, orange] (5.6,7) -- (5.6,8);
    \node[text=orange] at (5.6, 8.3) {\textbf{\(1.3 \)}};
    
    \draw[thick, orange] (5.6,4) -- (5.6,4.5);
    \draw[thick, orange] (5.6,1) -- (5.6,1.5);

    \node[left] at (0,7) {\textbf{$d_L$ [pc]}};
    \node[left] at (0,4) {\textbf{$f$ [Hz]}};
    \node[left] at (0,1) {\textbf{$T$ [yr]}};

    \node[text=bandColor] at (2,6.2) {\textcolor{bandColor}{\(3.1\cdot10^{-3} \) }};
    \node[text=bandColor] at (5,6.2) {\textcolor{bandColor}{\(5.1 \cdot 10^{-1}\) }};
    
    \node[text=bandColor] at (2,3.2) {\textcolor{bandColor}{\(5 \cdot 10^{-7}\) }};
    \node[text=bandColor] at (5,3.2) {\textcolor{bandColor}{\(3 \cdot 10^{-9}\) }};
    
    \node[text=bandColor] at (2,0.2) {\textcolor{bandColor}{\(\sim 20 \) days }};
    \node[text=bandColor] at (5,0.2) {\textcolor{bandColor}{\(10.5\) yr}};

    \node[text=pinkish] at (4.9,1.3) {\textcolor{pinkish}{\(\sim 15\)}};

    \node[text=bandColor] at (5, 4.4) {\textbf{\(\frac{1}{T}\)}};
    \node[text=bandColor] at (2, 4.4) {\textbf{\(\frac{1}{\Delta t}\)}};
    
    \filldraw[blueColor, opacity=0.5] (6.3,7) rectangle (8.3,7.2);
    \filldraw[blueColor, opacity=0.5] (6.3,4) rectangle (8.3,4.2);
    \filldraw[blueColor, opacity=0.5] (6.3,1) rectangle (8.3,1.2);  
    
    \node[white] at (7.2, 7.1) {\tiny \textbf{GAIA Peak}};
    \node[text=gaiaColor] at (8.3, 6.7) {\textbf{\(10^4\)}};
    \node[text=gaiaColor] at (6.3, 6.7) {\textbf{\(10^2\)}};

    \node[text=gaiaColor] at (8.4, 3.7) {\textbf{\(1.5\cdot 10^{-13}\)}};
    \node[text=gaiaColor] at (6.2, 3.7) {\textbf{\(1.5\cdot 10^{-11}\)}};

    \node[text=gaiaColor] at (8.5, 0.7) {{\(0.203 \) Myr}};
    \node[text=gaiaColor] at (6.3, 0.7) {{\(2038 \)}};
    
    \node[text=lightgreen] at (-0.2,6.2) {\textcolor{lightgreen}{\(2.8 \cdot 10^{-6}\)}};
    \node[text=lightgreen] at (-0.2,3.2) {\textcolor{lightgreen}{\(5.6 \cdot 10^{-4}\)}};
    \node[text=lightgreen] at (-0.2,0.2) {\textcolor{lightgreen}{\(\sim 30\) min}};

    \node[text=lightgreen] at (4, 8.1) {\textbf{\(2\cdot10^{-2} \)}};
    \draw[thick, lightgreen] (3.5,7) -- (3.5,7.7);  
    \node[text=lightgreen] at (3, 5) {\textbf{\(7.7 \cdot 10^{-8} \)}};
    \draw[thick, lightgreen] (3.5,4) -- (3.5,4.7);
    \node[text=lightgreen] at (3, 2) {{\(5\) months}};
    \draw[thick, lightgreen] (3.5,1) -- (3.5,1.7);
    
    \node[text=purple] at (0.45, 9.4) {Jupyter Orbit};
    \node[text=purple] at (0.45, 9.0) {\textbf{\(2.5 \cdot 10^{-5} \)}};
    \draw[thick, purple] (0.45,7) -- (0.45,8.7);      
    
    \node[text=brown] at (-1.2, 8.2) {{Asteroid Belt}};
    \node[text=brown] at (-1.2, 7.7) {\textbf{\(1.5 \cdot 10^{-5} \)}};
    \draw[thick, brown] (0.35,7) -- (0.35,8); 
    
    \node[text=cyan] at (2.1, 8.5) {Kuiper Belt};
    \node[text=cyan] at (2.3, 8.1) {\textbf{\(2 \cdot 10^{-4} \)}};
    \draw[thick, cyan] (1.5,7) -- (1.5,8);

    \node[text=magenta] at (8.5, 8.4) {\textbf{Milky Way}};
    \node[text=magenta] at (8.5, 8) {\textbf{\(3.2 \cdot 10^4\)}};
    \draw[thick, magenta] (8.5,7) -- (8.5,7.7);

    \node[text=magenta] at (8.5, 5) {\textbf{\(4.8 \cdot 10^{-14}\)}};
    \draw[thick, magenta] (8.5,4) -- (8.5,4.7);

    \node[text=magenta] at (8.5, 2) {{\(0.652 \) Myr}};
    \draw[thick, magenta] (8.5,1) -- (8.5,1.7);

    
    \node[text=darkgreen] at (14, 8.6) {\textbf{Universe Age}};
    \node[text=darkgreen] at (14, 8.1) {\textbf{\(6.73 \cdot 10^{8}\)}};
    \draw[thick, darkgreen] (14,7) -- (14,7.7);

    \node[text=darkgreen] at (14,5) {\textbf{\(2.3 \cdot 10^{-18}\)}};
    \draw[thick, darkgreen] (14,4) -- (14,4.7);

    \node[text=darkgreen] at (14, 1.9) {{\(13.8 \) Gyr}};
    \draw[thick, darkgreen] (14,1) -- (14,1.7);    

    \foreach \x/\dist/\freq/\T in {
        1/{1.55 \cdot 10^{-4}}/{ 10^{-5}}/{1.1 \, \text{days}}, 
        4/{1.55 \cdot 10^{-1}}/{ 10^{-8}}/{3.17}, 
        7/{1.55 \cdot 10^{2}}/{ 10^{-11}}/{3.17 \cdot 10^3}, 
        10/{1.55 \cdot 10^{5}}/{ 10^{-14}}/{3.17 \cdot 10^6}, 
        13/{1.55 \cdot 10^{8}}/{ 10^{-17}}/{3.17 \cdot 10^9}, 
        16/{1.55 \cdot 10^{11}}/{ 10^{-20}}/{3.17 \cdot 10^{12}}} {
        
        \draw[thick] (\x,7.2) -- (\x,6.8); 
        \draw[thick] (\x,4.2) -- (\x,3.8); 
        \draw[thick] (\x,1.2) -- (\x,0.8); 
       
        \node at (\x,7.5) {\(\dist\)}; 
        \node at (\x,4.5) {\(\freq\)}; 
        \node at (\x,1.5) {\(\T\)}; 
    }
\end{tikzpicture}
}

\end{minipage}
\caption{\emph{Top panel}: Regions in the~$d_L-f$ plane indicating current and future surveys sensitivities to a GW background together with the source's distances probed by different survey. 
The lines corresponding to~$\tilde{x}=10^{-1},1,10$ indicate the short-distance limit, the transition region, and the region where the long-distance approximation starts to apply, respectively. 
\emph{Bottom panel}: Mapping between frequencies, sources' distances and observation times. 
The correspondence is made in a way that given a frequency, the corresponding distance returns $\tilde{x}=1$.}
\label{fig::sources_dete}
\end{figure}
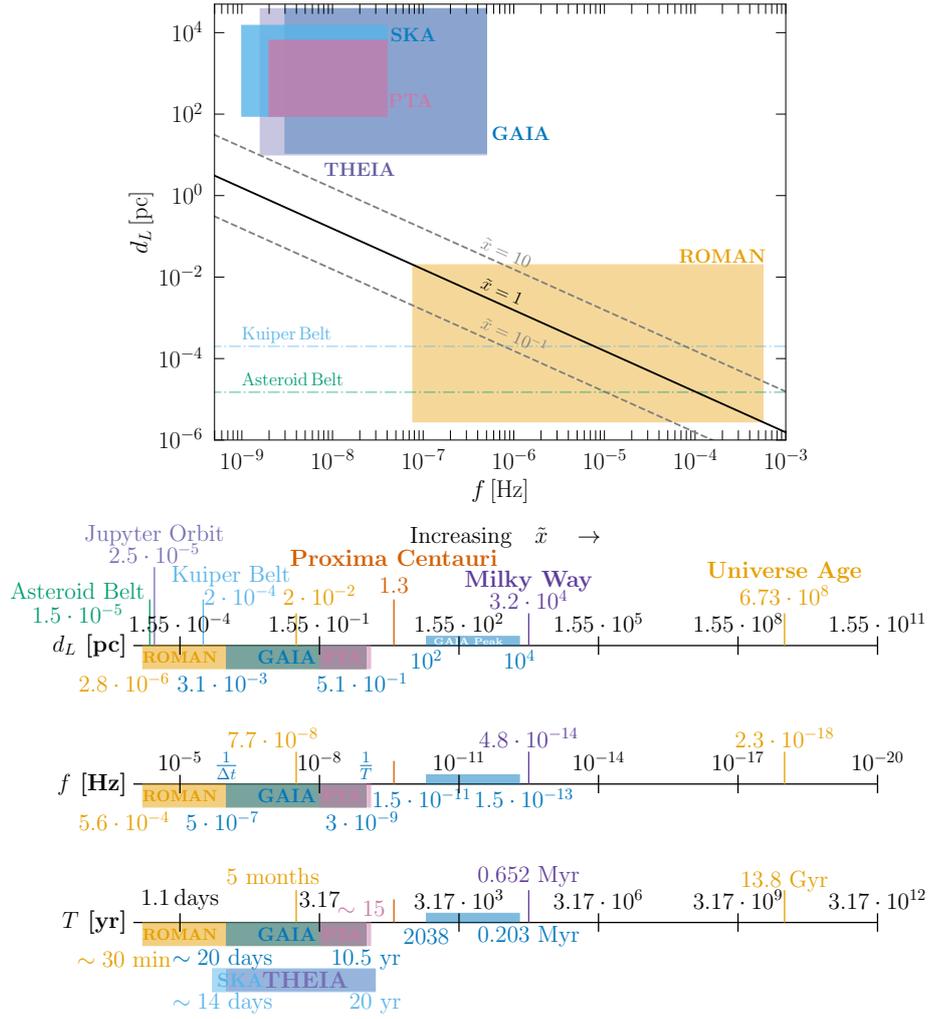

We provide in figure~\ref{fig::sources_dete} an intuitive graphical representation that illustrates the correspondence between current and future detectors and the characteristic distances and frequency ranges of the sources they can observe.
In the top panel of the figure we show the potential GW frequency sensitivity bands of different experiments, estimated as the interval between the inverse of the total observation time and the inverse of the cadence.
The figure also shows lines corresponding to~$\tilde{x}=10^{-1}, 1, 10$, which provide an intuitive reference for the regime  where the infinite-distance limit applies. 
On the other hand, in the bottom panel of the same figure we report a visual compact way to relate to each frequency/distance the corresponding distance/frequency that returns $\tilde{x}=1$, along with the observation time needed to observe a specific frequency and the bands associated to the frequency range available to each detector. 
This panel is read as follows.
First, we choose a frequency~$f_p$ on the second line.
The distance~$d_L^p$ which is vertically aligned with the chosen frequency returns~$\tilde{x}=1$. 
All sources at distances~$d_L>d_L^p$ will return~$\tilde{x}>1$, while sources at~$d_L<d_L^p$ correspond to~$\tilde{x}<1$. 
The third line, instead, shows the observational time necessary to probe those frequencies.
Similarly, after fixing a distance and recovering the corresponding frequency that gives~$\tilde{x}=1$, we have that all frequencies larger (smaller) than that correspond to~$\tilde{x}>1$ ($\tilde{x}<1$). 
As an example, in the case of GAIA we read from the second line that the corresponding frequency range would be roughly in the interval~$[3\cdot10^{-9},5\cdot10^{-7}]$ Hz, which gives $\tilde{x}\sim 1$ or lower only if the targeted sources are in the range $d_L\in[5.1\cdot10^{-5},3.1\cdot10^{-3}]$ pc. 
However, the peak of sources observed by GAIA is in the range $d_L\in[10^2,10^{4}]$ pc, thus in the long-distance limit. 
In the case of Roman, instead, it is easy to conclude that monitoring asteroids in the solar system would necessarily require the full analysis that includes the distance form factor.

In this work, we rely on a harmonic decomposition to estimate the importance of a synergic effort between astrometric and pulsar timing observations.
Since the harmonic power spectra, the $C_\ell$s, are obtained by integrating the frequencies and the solid angle, in the following, we find it convenient to work with the frequency domain angular displacement
\begin{equation}
\label{eq::dn_full}
    \begin{aligned}
        \delta n^i (f,r,\hat{\mathbf{n}}) &= \sum_\lambda \int d\hat{\mathbf{p}}\ h_\lambda(f,\hat{\mathbf{p}}) \left[ \frac{p^i + n^i}{2(1+\hat{\mathbf{p}}\cdot\hat{\mathbf{n}})} n^j n^k e^\lambda_{jk}(\hat{\mathbf{p}}) - \frac{1}{2} n^j e^\lambda_{ij}(\hat{\mathbf{p}}) \right. \\
        &\qquad\qquad \left. + i\mathcal{D} (f,\mathbf{r}) \left( \frac{p^i n^j n^k e^\lambda_{jk}(\hat{\mathbf{p}})}{2(1+\hat{\mathbf{p}}\cdot\hat{\mathbf{n}})} - n^j e^\lambda_{ij}(\hat{\mathbf{p}}) + \frac{(2+\hat{\mathbf{p}}\cdot\hat{\mathbf{n}})}{2(1+\hat{\mathbf{p}}\cdot\hat{\mathbf{n}})} n^i n^j n^k e^\lambda_{jk}(\hat{\mathbf{p}}) \right) \right],
    \end{aligned}
\end{equation}
which, given the vectorial nature of the deflection, we further decompose as
\begin{equation}
    \delta \mathbf{n} (f,r,\hat{\mathbf{n}}) = \sum_{\ell m} \left[ E^{f,r}_{\ell m} \mathbf{Y}^E_{\ell m} (\hat{\mathbf{n}}) + B^{f,r}_{\ell m} \mathbf{Y}^B_{\ell m} (\hat{\mathbf{n}}) \right]\,.
\end{equation}
Here~$\mathbf{Y}^E_{\ell m}, \mathbf{Y}^B_{\ell m}$ are vector spherical harmonics and~$E^{f,r}_{\ell m},B^{f,r}_{\ell m}$ are frequency- and distance-dependent harmonic coefficients.
The explicit definitions of vector spherical harmonics are reported in appendix~\ref{app:notation_conventions}, while in appendix~\ref{app:explicit_Cl_calculation} we report the definition of the E- and B-mode harmonic coefficients.
The expectation value of the E and B coefficients reads
\begin{equation}
    \begin{aligned}
        \left\langle  E^{f,r}_{\ell m} \left(E^{f',r'}_{\ell m}\right)^* \right\rangle &= \frac{1}{2} \delta^D(f-f') \delta^K_{\ell\ell'} \delta^K_{mm'} C^{EE}_\ell(f,r,r')\,,\\
        \left\langle  B^{f,r}_{\ell m} \left(B^{f',r'}_{\ell m}\right)^* \right\rangle &= \frac{1}{2} \delta^D(f-f') \delta^K_{\ell\ell'} \delta^K_{mm'} C^{BB}_\ell(f,r,r')\,,
    \end{aligned}
\end{equation}
where the angular power spectra result
\begin{equation}
    \begin{aligned}
        C^{EE}_\ell (f,r,r') &= \frac{8\pi}{\ell(\ell+1)} \frac{(\ell-2)!}{(\ell+2)!} F^E_\ell(2\pi fr) \left[F^E_\ell(2\pi f r')\right]^* S_h(f)\,, \\
        C^{BB}_\ell (f,r,r') &= \frac{8\pi}{\ell(\ell+1)} \frac{(\ell-2)!}{(\ell+2)!} F^B_\ell(2\pi fr) \left[F^B_\ell(2\pi f r')\right]^* S_h(f)\,.
    \end{aligned}
\end{equation}
The derivation of this result is reported in appendix~\ref{app:explicit_Cl_calculation}.
Their frequency-integrated counterpart are
\begin{equation}
\label{Eq::Clrrp}
    \begin{aligned}
        C^{EE}_\ell (r,r')  &= \int_{-\infty}^\infty df df' \left\langle  E^{f,r}_{\ell m} \left(E^{f',r'}_{\ell m}\right)^* \right\rangle \\
        &= \frac{8\pi}{\ell(\ell+1)} \frac{(\ell-2)!}{(\ell+2)!} \int_{0}^{+\infty} df\ \mathrm{Re}\left[F^E_\ell (2\pi fr) \left[F^E_\ell(2\pi f r')\right]^*\right] S_h(f)\,, \\
        C^{BB}_\ell (r,r')  &= \int_{-\infty}^\infty df df' \left\langle  B^{f,r}_{\ell m} \left(B^{f',r'}_{\ell m}\right)^* \right\rangle \\
        &= \frac{8\pi}{\ell(\ell+1)} \frac{(\ell-2)!}{(\ell+2)!} \int_{0}^{+\infty} df\ \mathrm{Re}\left[F^B_\ell (2\pi fr) \left[F^B_\ell(2\pi f r')\right]^*\right] S_h(f)\,. \\
    \end{aligned}
\end{equation}
These last expressions already introduce the novelty of this work. 
Contrary to previous analyses, these angular power spectra include the dependence on the distances of the sources.
One of the main results of this paper, explicitly derived in appendix~\ref{app:explicit_Cl_calculation}, is the calculation of the E and B form factors
\begin{equation}
\label{Eq:ClBB_ClEE}
    \begin{aligned}
        F^E_\ell (\tilde{x}) &= i^\ell e^{i\tilde{x}} \left[ j_{\ell-1}(\tilde{x}) \left(\frac{\ell(\ell+1)}{2\tilde{x}} + \tilde{x} \right) - j_\ell(\tilde{x}) \left( i\tilde{x} - \frac{\ell^2(\ell^2-1)}{4\tilde{x}^2} + \ell \right)\right], \\
        F^B_\ell (\tilde{x}) &=(-i) i^\ell e^{i\tilde{x}} \left[ j_\ell(\tilde{x}) \left(i \frac{\ell(\ell-1)}{2} + \frac{\ell(\ell^2-1)}{2\tilde{x}} + \tilde{x} \right) + j_{\ell-1}(\tilde{x}) \left( i\tilde{x} - \frac{\ell(\ell+1)}{2} \right) \right], \\
    \end{aligned}
\end{equation}
that precisely arise from the finite distance contribution coming from the term proportional to~$\mathcal{D}(f,\mathbf{r})$.

\begin{figure}[ht]
    \centerline{
    \includegraphics[width=\linewidth]{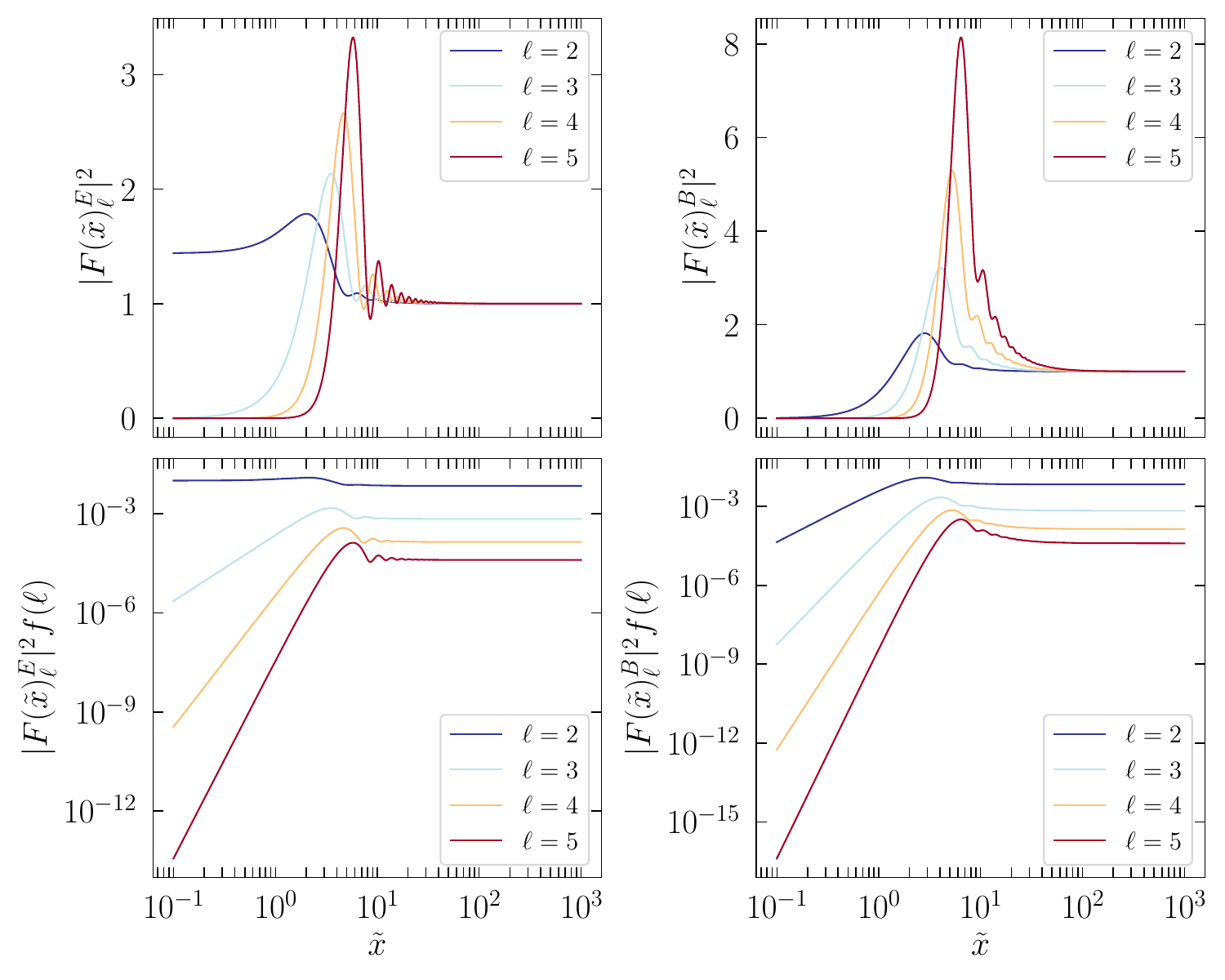}}
    \caption{\emph{Top panels}: form factors for the E- and B-mode as function of~$\tilde{x}=2\pi f r$ for different values of~$\ell$. 
    \emph{Bottom panels}: behavior of the~$\ell$-dependent terms in equation~\eqref{Eq:ClBB_ClEE}. 
    Contrary to the top panel, the curves show that the first multipole gives the highest contribution for any value of~$\tilde{x}$.}
    \label{fig:FEB2_of_x}
\end{figure}

We show in figure~\ref{fig:FEB2_of_x} the absolute value of the~$F^E_\ell, F^B_\ell$ form factors as a function of~$\tilde{x}$ for the first four multipoles.
The importance of our approach is highlighted by the top panels of the figure, where we observe that these E- and B- form factors present a non-trivial pattern of oscillation and a distinctive behavior in the limit of small dimensionless~$\tilde{x}\to 0$.
In particular, we note that both these form factors present a rather sharp peak in the range of~$\tilde{x}\sim\mathcal{O}(1-10)$, and they all quickly vanish for~$\tilde{x}\lesssim 1$, except for~$F^E_2$ which tends to the constant value of~$|F_2^E|^2 \to 36/25$\footnote{
In the limit of small $\tilde{x}$, the E-mode form factor is proportional to~$F_\ell^E\propto \tilde{x}^{2\ell-4}$, that goes to zero except when~$\ell=2$. 
Conversely, $F_\ell^B\propto \tilde{x}^{2\ell-2}$; thus it vanishes for any $\ell\geq2$.}.
Finally, we observe that, for~$\ell=2$, the infinite distance limit for the E- and B-modes is recovered only when~$\tilde{x}\gtrsim 10$; whereas larger values of the multipole require~$\tilde{x}\geq 30$ and~$\tilde{x}\geq 50$ for the E- and B-mode form factor to converge to the infinite distance limit, respectively.
Thus, given a frequency band, some sources could satisfy the long distance approximation only for some multipoles, while still needing a full approach for others. 
In order to appreciate the full dependence on the multipole, we report in the lower panels of the figure the form factors weighted by the $\ell$-dependent terms present in equation~\eqref{Eq::Clrrp}, defined as~$f(\ell) = \frac{1}{\ell(\ell+1)} \frac{(\ell-2)!}{(\ell+2)!}$. 
These plots show that, for both E- and B-modes, the dominant contribution always comes from the first multipole, while high multipoles become progressively subdominant, for all values of $\tilde{x}$.

\begin{figure}[ht]
    \centerline{
    \includegraphics[width=0.9\linewidth]{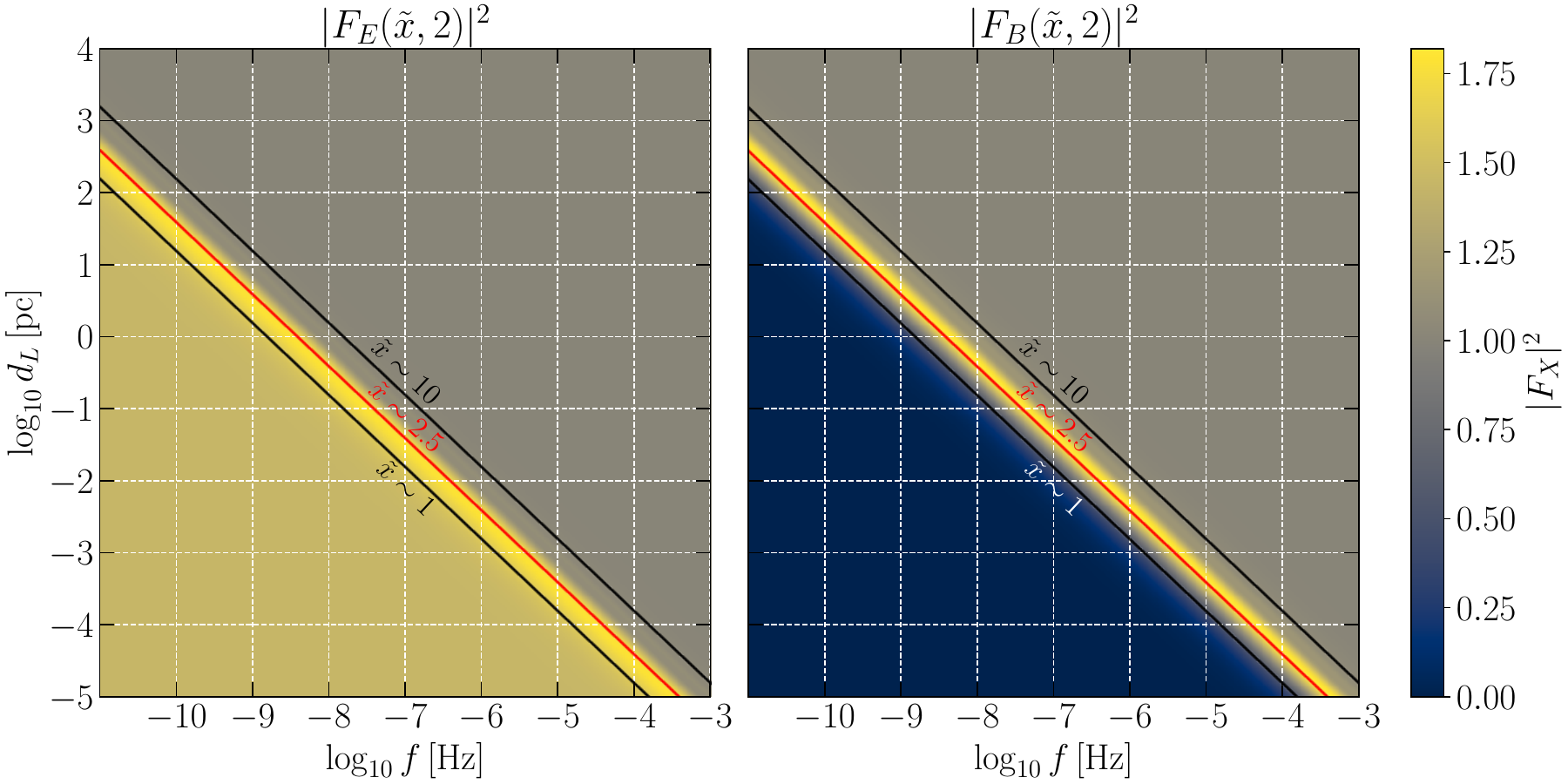}}
    \caption{Absolute value of the form factor for the E- (\textit{left panel}) and B-mode (\textit{right panel}) for different values of the frequency and distance. 
    Given a frequency one can easily understand if for a given source the long distance approximation holds and the corresponding value of the form factor.}
    \label{fig:FEB2_of_f_r}
\end{figure}
    
To better grasp where each observable lives in terms of distance and frequency probed, we show in figure~\ref{fig:FEB2_of_f_r} the value of E- and B-mode form factors for~$\ell=2$ as a function of both distances and GW frequency. 
We also show, as a reference, the lines corresponding to~$\tilde{x}=1$, i.e., where the wavelength of the GW is comparable to the distance of the monitored source, to~$\tilde{x}=2.5$, where the source term peaks, and to~$\tilde{x}=10$, where the long distance limit starts to be valid.
The figure highlights the importance of employing the full form factors, for instance when deflections for sources at distances of~$d_L \approx 10^{-4}-10^{-5}$ pc, like asteroids in the Solar System, in the presence of a GWB at frequencies of~$f \approx 10^{-3}-10^{-4}$ Hz. 
In other words, a GWB that could be probed by LISA has also the potential of imprinting a unique effect on the measured angular position of asteroids, highlighting the importance of a synergistic effort between astrometric probes and interferometric searches.


\subsection{Pulsar timing residual measurements}

Millisecond pulsars are characterized by a very stable rotational period~$T_\mathrm{psr}$.
However, the presence of a GWB modifies the geodesics of the emitted photons, and it is responsible for an additional delay~$\Delta T_\mathrm{GWB}$ in the time-of-arrival (TOA) of pulses.
In other words, the inferred period is given by
\begin{equation}
    T^\mathrm{obs}_\mathrm{psr} = \mathrm{TOA}_{j} - \mathrm{TOA}_{j-1} = T_\mathrm{psr} + \Delta T_\mathrm{GWB},
\end{equation}
where the index~$j$ labels the pulses, on top of the other astrophysical effects that are commonly implemented in a timing model, see, e.g., ref.~\cite{NANOGrav:2023hde}.
Therefore, by repeatedly measuring the TOA of pulses, we can infer the presence and properties of an intervening GWB.

For each pulsar at distance~$\mathbf{r}=r\hat{\mathbf{n}}$ from Earth, we define as the primary observable of the GWB the ``redshift'' variable
\begin{equation}
    z(t,\mathbf{r}) = \frac{\Delta T_\mathrm{GWB}}{T_\mathrm{psr}} = \sum_\lambda \int df d\hat{\mathbf{p}}\ h_\lambda(f,\hat{\mathbf{p}}) \frac{n^i n^j e^\lambda_{ij}(\hat{\mathbf{p}})}{2(1+\hat{\mathbf{n}} \cdot \hat{\mathbf{p}})} e^{-2\pi i ft} \left[1 - e^{2\pi i f r(1 + \hat{\mathbf{p}} \cdot \hat{\mathbf{n}})} \right].
\label{eq:redshift_timing_residuals}
\end{equation}
In reality, all galactic pulsars that are typically included in a PTA analysis are at distances of order~$r\sim\mathcal{O}(\mathrm{kpc})$, i.e., in the so-called ``long-arm limit'' for a nHz GWB, i.e.,~$2\pi fr \approx 10^3\gg 1$.
In this scenario, the content of square brackets in equation~\eqref{eq:redshift_timing_residuals} can be safely neglected because its rapidly oscillatory contribution averages out.
Therefore, we have that the redshift observable in the frequency domain is well-approximated by
\begin{equation}
    z(f,\hat{\mathbf{n}}) = \sum_\lambda \int d\hat{\mathbf{p}}\ h_\lambda(f,\hat{\mathbf{p}}) F_\lambda(\hat{\mathbf{n}}, \hat{\mathbf{p}})\,,
\end{equation}
where the antenna pattern function reads as
\begin{equation}
    F_\lambda(\hat{\mathbf{n}}, \hat{\mathbf{p}}) = \frac{n^i n^j e^\lambda_{ij}(\hat{\mathbf{p}})}{2(1+\hat{\mathbf{n}} \cdot \hat{\mathbf{p}})}\,.
\end{equation}
It immediately descends from the GWB statistical properties that the redshift two-point function is given by
\begin{equation}
    \begin{aligned}
        \left\langle z(f,\hat{\mathbf{n}}) z^*(f',\hat{\mathbf{n}}') \right\rangle &= \frac{1}{2}\delta^D(f-f') S_h(f) \mathrm{HD}\,, \\        
    \end{aligned}
\end{equation}
where~$\mathrm{HD}(\cos\theta)$ represents the Hellings-Downs curve~\cite{Hellings:1983fr}, explicitly defined as
\begin{equation}
    \mathrm{HD}(\hat{\mathbf{n}}\cdot\hat{\mathbf{n}}') = \int \frac{d\hat{\mathbf{p}}}{4\pi} \sum_\lambda F_\lambda(\hat{\mathbf{n}}, \hat{\mathbf{p}}) F_\lambda(\hat{\mathbf{n}}', \hat{\mathbf{p}}) = \frac{1}{3} - \frac{1-\hat{\mathbf{n}}\cdot\hat{\mathbf{n}}'}{12} + \frac{1-\hat{\mathbf{n}}\cdot\hat{\mathbf{n}}'}{2}\log \frac{1-\hat{\mathbf{n}}\cdot\hat{\mathbf{n}}'}{2}.
\end{equation}
By decomposing the redshift in spherical harmonics as
\begin{equation}
    z(f,\hat{\mathbf{n}}) = \sum_{\ell m} z^f_{\ell m} Y_{\ell m} (\hat{\mathbf{n}}),
\end{equation}
where~$Y_{\ell m}$ are the spherical harmonics and~$z^f_{\ell m}$ are the frequency-dependent harmonic coefficients, we obtain that their two-point function reads as
\begin{equation}
    \left\langle z^f_{\ell m} z^{f'}_{\ell' m'} \right\rangle = \frac{1}{2}\delta^D(f-f') \delta^K_{\ell\ell'} \delta^K_{mm'} C^{zz}_\ell(f),
\end{equation}
where the angular power spectrum is
\begin{equation}
    C^{zz}_\ell(f) = 4\pi \frac{(\ell-2)!}{(\ell+2)!} S_h(f),
\end{equation}
or, in a more realistic scenario, where we integrate the two-point function over frequencies, we have
\begin{equation}
    C_\ell^{zz}=\int df df' \left\langle z^f_{\ell m} z^{f'}_{\ell' m'} \right\rangle = \delta^K_{\ell\ell'} \delta^K_{mm'} C^{zz}_\ell = \delta^K_{\ell\ell'} \delta^K_{mm'} \times 4\pi \frac{(\ell-2)!}{(\ell+2)!} \int df S_h(f).
\end{equation}
Also in this case, the interested reader can find explicit definitions and properties of spherical harmonics in appendix~\ref{app:notation_conventions}, and the full derivation of the angular power spectrum in appendix~\ref{app:explicit_Cl_calculation}.


\section{Observational targets}
\label{sec:observational_targets}


\subsection{Compressed statistics}

In the previous section, we have characterized both the time delays and the apparent displacement of individual astrophysical objects.
However, in order to process any vast amount of data, as in the case at hand, we are forced to work with compressed statistics, especially in the context of astrometric probes, where the number of individual sources ranges from millions to billions.

In this sense, for objects not in the infinite distance limit, it is convenient to introduce a ``volume-averaged'' displacement to describe the statistics of the population instead of focusing on individual objects.
Given a finite volume element~$V$ of a spherical shell at distance~$\bar{r}$ and with half-width~$\Delta r$ containing~$N_\mathrm{obj}$ objects, we define the average displacement as
\begin{equation}
    \delta \mathbf{n} (f, \bar{r}, \hat{\mathbf{n}}) = \frac{\sum_{j\in V} \delta \mathbf{n} (f,r_j,\hat{\mathbf{n}})}{N_\mathrm{obj}} = \frac{1}{N(\bar{r},\Delta r)}\int _0^\infty dr \frac{dN_\mathrm{obj}}{dr} W_r(r, \bar{r}, \Delta r) \delta \mathbf{n} (f,r,\hat{\mathbf{n}}) \,,
\end{equation}
where the RHS represents the description in the continuum limit, $dN_\mathrm{obj}/dr$ is the radial number density distribution, and~$W_r$ is the window function defining the distance and thickness of the spherical shell. 
The function~$N(\bar{r},\Delta r)$ is a normalization constant defined as
\begin{equation}
    N(\bar{r},\Delta r) = \int _0^\infty dr \frac{dN_\mathrm{obj}}{dr} W_r(r, \bar{r}, \Delta r)  \,.
\end{equation}

On the practical side, the quantity that is going to appear in the angular power spectra is a distance-averaged form factor
\begin{equation}
    F^X_\ell(f, \bar{r}) = \frac{1}{N(\bar{r},\Delta r)} \int_0^\infty dr \frac{dN_\mathrm{obj}}{dr} W_r(r, \bar{r}, \Delta r) F^X_\ell (2\pi f r).
\end{equation}
In the following, we implement a number density radial distribution of the form
\begin{equation}
    \frac{dN_\mathrm{obj}}{dr} = N^\mathrm{tot}_\mathrm{obj}  \Phi_\mathrm{obj}(r),
\end{equation}
where~$N^\mathrm{tot}_\mathrm{obj}$ is the total number of objects of our sample between~$[r_\mathrm{min},r_\mathrm{max}]$, i.e., between the minimum and maximum distance of the objects from the observer, and~$\Phi_\mathrm{obj}$ is a probability distribution that describes the shape of the radial distribution and is normalized to unity, i.e.,~$\displaystyle \int_{r_\mathrm{min}}^{r_\mathrm{max}} dr \Phi_\mathrm{obj}(r) = 1$.
Additionally, when the uncertainty on the distance is typically much smaller than the distance itself, a reasonable choice of window function is a top-hat function of the form
\begin{equation}
    W_r(r, \bar{r}, \Delta r) = \Theta_H \left(\bar{r} + \Delta r - r \right) \Theta_H \left(r - \bar{r} + \Delta r \right).
\end{equation}

Finally, we note that no individual experiment is sensitive to the entire range of GW frequencies, especially in light of the fact that none of the measurements occurs on a continuous basis in time.
To account for this effect, we introduce a window function in frequency-space~$W_f$, so that for each observable~$X(f,\mathbf{r})$ what we actually compare with data is
\begin{equation}
    X(\bar{f},\mathbf{r}) = \int df W_f(f, \bar{f}, \Delta f) X(f,\mathbf{r}),
\end{equation}
where~$\bar{f}$ and~$\Delta f$ are the representative average frequency and width of the chosen frequency bin and, in this case, the window function does not need to be normalized to unity, since we want to consider the integrated effect of all frequencies.
The most simple and conservative choice is represented by
\begin{equation}
    W_f(f, \bar{f}, \Delta f) = \Theta_H \left(f - \bar{f} + \Delta f\right) \Theta_H \left(\Delta \bar{f} + \Delta f - f\right) = \Theta_H \left(f-f_\mathrm{min}\right) \Theta_H \left(f_\mathrm{max} - f \right),
\end{equation}
where the minimum~$f_\mathrm{min}=T^{-1}_\mathrm{obs}$ and maximum~$f_\mathrm{max}=\Delta t^{-1}$ frequencies correspond to the inverse of the total observation time and cadence of observations, respectively.
Other window choices are possible; in particular, it might be convenient to introduce multiple frequency bin to probe the shape of the GWB power spectrum and to remove noisy frequency channel from the analysis.

In the end, the final form of the angular displacement power spectra is
\begin{equation}
    C^{X_iX_j}_\ell = \frac{8\pi}{\ell(\ell+1)} \frac{(\ell-2)!}{(\ell+2)!} \int df \mathrm{Re} \left[ F^X_\ell(f, \bar{r}_i) \left[F^X_\ell(f, \bar{r}_j) \right]^* \right] W_f^iW_f^j S_h(f)\,.
\end{equation}
When $i=j$, one straightforwardly obtains~$W_f^iW_f^i=|W_f|^2$.
However, for different sources (and eventually different detectors), as in the case of cross-correlations, only those frequencies that lie in the overlapping region of the two detectors should be considered, otherwise the integral will vanish. 
For the redshift angular power spectra we have
\begin{equation}
\label{eq::Cl_zz}
    C^{zz}_\ell = 4\pi \frac{(\ell-2)!}{(\ell+2)!} \int df |W_f|^2 S_h(f)\,.
\end{equation}


\subsection{Milky Way stars}

During the course of~$T_\mathrm{obs} = 5\ \mathrm{yrs}$, the GAIA satellite has measured the positions of about $N_\mathrm{star} = 1.9 \times 10^9$ stars, with a cadence of approximately~$\Delta t = 20\ \mathrm{days}$, providing an unprecedented high-resolution map of the Milky Way.
In this work, we aim at taking full advantage of the 3D distribution of stars, thus we divide the sample of stars into three subsamples, depending on their distance from us.
Parallaxes alone are not sufficient to provide an accurate estimate of star distance; however, Bayesian methods have been developed to incorporate into the analysis also the information coming from other star properties (color, magnitude, etc.) to provide a more robust estimate of star distances.
Therefore, in the following, we use the public catalog of distances obtained with such a method for the GAIA EDR3 sample~\cite{Bailer_Jones_2021}\footnote{The catalog is available in the public repository at \url{https://gea.esac.esa.int/archive/}.}.

\begin{table}[ht]
    \centerline{
    \begin{tabular}{|c|c|c|c|}
        \hline
        Family & CS & IS & FS \\
        \hline
        \hline
        $r_\mathrm{med}$ [pc] & $[11, 316]$ & $[316, 3160]$ & $[3160, 38600]$ \\
        $N_\mathrm{star}$ & $11.4 \times 10^6$ & $623 \times 10^6$ & $1.27 \times 10^9$ \\
        $\sigma_r$ [pc] & $32$ & $360$ & $1500$ \\
        $\sigma_\theta$ [mas] & $0.3$ & $0.3$ & $0.6$ \\
        \hline
        \hline
        $\bar{r}$ [pc] & $163.5$ & $1738$ & $20880$ \\
        $\Delta r$ [pc] & $152.5$ & $1422$ & $17720$ \\
        \hline
    \end{tabular}}
    \caption{The top part of the table contains the typical parameters of the three families of star populations analyzed in this work, whereas the bottom part contain the values that define the radial window function for each family.}
    \label{tab:gaia_stars}
\end{table}

We divide our sample of stars on the basis of their Bayesian median distance~$r_\mathrm{med}$ into three distance bins of 
close-by (CS), intermediate (IS), and far-away (FS) stars.
The different distance bins and the total number of stars for each bin are reported in table~\ref{tab:gaia_stars}, together with the estimate of their average errors on distance~$\sigma_r$ and angular position~$\sigma_\theta$.
Additionally, we report in appendix~\ref{app:population_properties} the complete probability distribution function for both radial and angular errors.

\begin{figure}[ht]
    \centerline{
    \includegraphics[width=\linewidth]{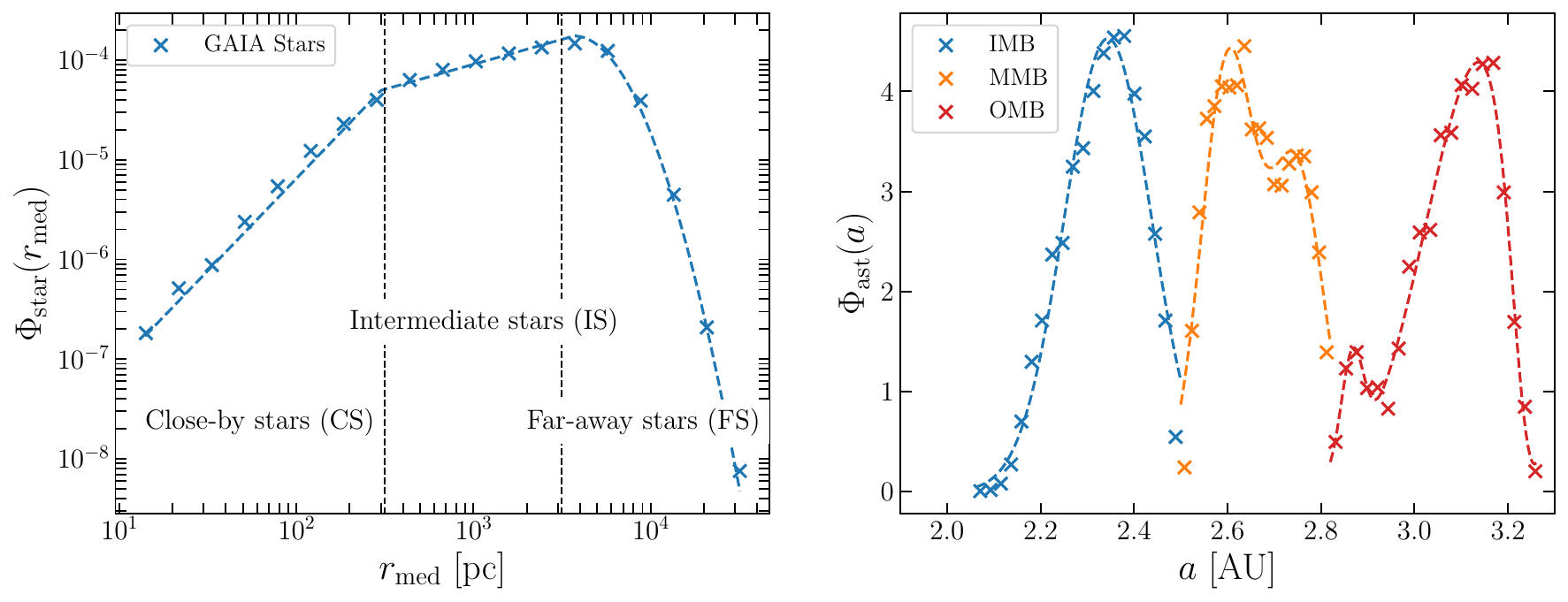}}
    \caption{Radial number density probability distribution for stars in the GAIA sample (\textit{left panel}) and Main Belt asteroids in the NASA Small-Body Database (\textit{right panel}).
    Markers indicate the value measured from the catalogs, and dashed lines represents the values of our fitting formula.}
    \label{fig:star_asteroid_populations}
\end{figure}

We show in the left panel of figure~\ref{fig:star_asteroid_populations} the radial probability distribution function for all stars of the sample. 
We report both the values measured from the catalogs and the profile described by our fitting formula~$\Phi(r_\mathrm{med})$, which we also explicitly report, together with the best-fit parameters, in appendix~\ref{app:population_properties}.
The rapid drop in distribution at distances greater than a few kpc is due to the fact that the Milky way does not have a symmetric structure around our position, and that stars behind the galactic center are hidden from our view.
In the following, we assume that~$\Phi_\mathrm{star}(r) = \Phi_\mathrm{star}(r_\mathrm{med})$.
In terms of sky coverage, it is not unrealistic to assume that stars are approximately uniformly distributed for what concerns the first two distance bins; and that, especially for very distant objects, the distribution becomes not uniform.

Finally, in the bottom part of table~\ref{tab:gaia_stars} we report the parameters that would describe a top-hat window function for the three radial bins.
Since for all bins the average distance error is smaller than the width of the distance bin, i.e., $\sigma_r \ll 2\Delta r$, we can safely conclude that a top-hat window function is a reasonable choice to implement.
This latter condition has also been explicitly using the error distance probability distribution function reported in appendix~\ref{app:population_properties}.
In particular, we find the probability that~$p(\sigma_r \geq 2\Delta r) <1\%$ for all three families of stars.


\subsection{Main Belt asteroids}

The Main Belt is populated by asteroids with semimajor axis~$a$ between~$[2.06, 3.27]\ \mathrm{AU}$ orbiting the Sun.
This class of asteroids is actually divided into three families, Inner Main Belt (IMB), Middle Main Belt (MMB), and Outer Main Belt (OMB), depending on their distance from the Sun.
The subdivision arises quite naturally because of the presence of the so-called Kirkwood gaps, which are created by orbital resonances with Jupiter.
In particular, the IMB, MMB, and OMB are situated between the~$(4:1-3:1)$, $(3:1-5:2)$, and~$(5:2-2:1)$ period ratio, respectively, although some of these families in reality present a substructure due to the presence of weaker resonances, as we can also see on the right panel of figure~\ref{fig:star_asteroid_populations}.

\begin{table}[ht]
    \centerline{
    \begin{tabular}{|c|c|c|c|}
        \hline
        Family & IMB & MMB & OMB \\
        \hline
        \hline
        $N_\mathrm{ast}$ & $364000$ & $500000$ & $458000$ \\
        $a$ [AU] & $[2.06, 2.5]$ & $[2.5, 2.82]$ & $[2.82, 3.27]$ \\
        $\sigma_{a}$ [AU] & $0.04$ & $0.02$ & $0.07$ \\
        \hline
        \hline
        $\bar{r}$ [AU] & $2.28$ & $2.66$ & $3.045$ \\
        $\Delta r$ [AU] & $0.22$ & $0.16$ & $0.225$ \\
        \hline
    \end{tabular}}
    \caption{The top part of the table contains the total number of asteroids~$N_\mathrm{ast}$, their semimajor axis~$a$ and its average error estimate~$\sigma_a$, whereas the bottom part contains the values of the top-hat window function center and half-width, $\bar{r}$ and~$\Delta r$, respectively.}
    \label{tab:main_belt_asteroids}
\end{table}
In this work, we use the NASA public Small-Body Database to recover the orbital parameters of IMB, MMB, and OMB asteroids, for a total of~$N^\mathrm{tot}_\mathrm{obj} \simeq 1.32\times 10^6$ objects with known orbital motion\footnote{
This database is part of the broader JPL Solar System Dynamics Database, available at \url{https://ssd.jpl.nasa.gov/}.}.
We report in table~\ref{tab:main_belt_asteroids} each family their total number of objects, semimajor axis and its error, and characteristic parameter of the window function.
We display the asteroid radial probability distribution in the right panel of figure~\ref{fig:star_asteroid_populations}, together with a fitting formula reported in appendix~\ref{app:population_properties}.
Although not reported in the table, we show in appendix~\ref{app:population_properties} that almost the totality of objects have eccentricities~$e \lesssim 0.3$; therefore, we can safely approximate the semimajor axis with the average radial distance from the Sun; therefore, in the following we use~$\Phi_\mathrm{ast}(r)=\Phi_\mathrm{ast}(a)$.
In this sense, the minimum and maximum distance of objects in the asteroid sample is~$r_\mathrm{min} = 2.06\ \mathrm{AU}$ and~$r_\mathrm{min} = 3.27\ \mathrm{AU}$, respectively.
Also in this case, we observe that a top-hat choice in the case of asteroids is appropriate since the average error on the semimajor axis is considerably smaller than the distance bin width, i.e.,~$\sigma_a \lesssim 2\Delta r$.
Finally, since most of the asteroids have relatively low inclinations, i.e., $i \lesssim 20^o$, as we also report in appendix~\ref{app:population_properties}, we have that these measurements have the potential to cover up to~$50\%$ of the sky.

 
\subsection{Galactic pulsars} 

The vast majority of currently known galactic pulsars that are used for pulsar timing measurements are situated at about kpc distance.
Therefore, their effective radial distribution is almost irrelevant in our analysis since they are in the long-arm limit.
Current PTA experiments are monitoring about~$N_\mathrm{psr}\approx 60-80$ pulsar, both in the Northern (mainly by the IPTA Collaboration) and the Southern (mainly by the MeerKAT Collaboration) hemisphere.
However, next generation experiments as the SKAO, have the potential to detect up to~$10^4$ new pulsars, including~$10^3$ millisecond pulsars~\cite{Janssen:2014dka,Keane:2014vja}.
The goal of such experiments is to monitor about~$N_\mathrm{psr}=200$ pulsars with a fortnightly cadence~($\Delta t = 14\ \mathrm{days}$).
Since these pulsars are for the most part situated in our galaxy, we can expect a sky coverage of about~$30\%$.


\section{Statistical analysis}
\label{sec:forecasts}


\subsection{Gravitational wave background models}
\label{subsec::GW_models}
We test our formalism by examining different GWB signals. We model the GW signal by means of the usually adopted power law formula, i.e.
\begin{align}
    \Omega_{\rm GW}(f) = \Omega_s\left(\frac{f}{f_s}\right)^\gamma
\end{align}
where~$f_s$ is the pivot frequency, $\Omega_s$ is the amplitude at the pivot scale, and~$\gamma$ is the spectral index.
The model considered are  
\begin{itemize}
    \item \underline{\textbf{PTA GW backgrounds.}} 
    Recent PTA experiments found evidence for a GWB for the first time and the exact source of the signal is still under debate. Similarly as in PTA analysis, we consider two representative backgrounds in the PTA frequency band. 
    For the first we adopt the median values obtained for a general model of the timing-residual power spectral density with variable power-law exponent,
    obtaining $\Omega_s = 1.12 \times 10^{-7}$, $f_s \simeq 3.17 \times 10^{-8}\,\mathrm{Hz}$ and $\gamma \simeq 1.8$~\cite{NANOGrav:2023gor}. 
    For the second we adopt the median values obtained assuming the background is attributed to super-massive black hole binaries (SMBHBs) in their inspiralling phase; following the latest PTA analyses we take 
    $\Omega_s = 1.5 \times 10^{-8}$, $f_s \simeq 3.17 \times 10^{-8}\,\mathrm{Hz}$ and $\gamma = 2/3$~\cite{NANOGrav:2023gor}.
    
    \item \underline{\textbf{Cosmological background.}} As an example, we consider the case of a cosmic string signal, given its broadness in frequency range. We choose, in order to get an illustrative example, the following reference values $\gamma = 0$ and $\Omega_s = 10^{-8}$.
\end{itemize}


\subsection{Statistical tools}
\label{sec::Statistical_Tools}
Given the potential of these probes, also used in combination with each other, here we are interested in exploring two aspects.
First, we aim at quantifying the detectability of the signal for different combinations of data sets.
Then, once we identify the most promising experiments, we want to provide a sensitivity forecast for the most important GWB models.

In terms of detectability, we define a cumulative Signal-to-Noise ratio (SNR) of the form
\begin{equation}
    \mathrm{SNR}^2 = f_\mathrm{sky} \sum_\ell \frac{2\ell+1}{2} \mathrm{Tr}\left[\mathcal{C}_\ell \left(\mathcal{C}_\ell+\mathcal{N}_\ell\right)^{-1} \mathcal{C}_\ell \left(\mathcal{C}_\ell+\mathcal{N}_\ell\right)^{-1} \right],
\label{eq:SNR2}
\end{equation}
where~$f_\mathrm{sky}$ is the fraction of the sky covered by the experiment, and~$\mathcal{C}_\ell,\, \mathcal{N}_\ell$ are covariance matrices for the signal and noise.\footnote{
This SNR$^2$ provides a reasonable estimate of what is the statistical significance of a signal detection, in the scenario where the signal has a Gaussian distribution. 
In other words, it gives an idea about how many~``$\sigma$'' away from the null signal~$a_{\ell m}=0/\mathcal{C}_\ell = 0$ we are.}
Regarding the signal covariance matrix, in its most general form it is constructed as a symmetric matrix with entries given by the angular power spectra and reads as
\begin{equation}
    \mathcal{C}_\ell = 
    \begin{pmatrix}
    & & C^{zz}_\ell & C^{zE_{(1,1)}}_\ell & \cdots & C^{zE_{(n,p)}}_\ell & C^{zB_{(1,1)}}_\ell & \cdots & C^{zB_{(n,p)}}_\ell \\
    & & & C^{E_{(1,1)} E_{(1,1)}}_\ell & \cdots & C^{E_{(1,1)} E_{(n,p)}}_\ell & C^{E_{(1,1)} B_{(1,1)}}_\ell & \cdots & C^{E_{(1,1)} B_{(n,p)}}_\ell \\
    & & & & \ddots & \vdots & \vdots & \ddots & \vdots \\
    & & & & & C^{E_{(n,p)} E_{(n,p)}}_\ell & C^{E_{(n,p)} B_{(1,1)}}_\ell & \cdots & C^{E_{(n,p)} B_{(n,p)}}_\ell \\
    & & & & & & C^{B_{(1,1)} B_{(1,1)}}_\ell & \cdots & C^{B_{(1,1)} B_{(n,p)}}_\ell \\
    & & & & & & & \ddots & \vdots \\
    & & & & & & & & C^{B_{(n,p)} B_{(n,p)}}_\ell \\
    \end{pmatrix}.
\end{equation}
Although cumbersome, this form of the covariance matrix allows the possibility of including redshift correlations, and~$n$ and~$p$ distance and frequency bins, respectively, for the angular displacement.
However, in many instances, the analysis is extremely simplified because of the properties of the background.
For example, when different polarization degrees of freedom are independent, we have~$C^{z B_{(k,l)}}_\ell = C^{E_{(i,j)} B_{(k,l)}}_\ell \equiv 0$ for all possible combinations of indices.
Additionally, if each GWB source emits in a single frequency bin during the entire course of observations, all frequency bins are fundamentally uncorrelated.\footnote{We add that PTA measurements suffer from frequency correlations which can lead to biased estimates of the GWB spectral parameters~\cite{Crisostomi:2025vue}. We neglect them for the purposes of this analysis.}
Finally, in the case of searches for broad-band GW signals, where the frequency ranges typical of redshift and angular displacement measurements do not overlap, also all $z-E$ cross-correlations vanish ($C^{zE_{(k,l)}}_\ell \equiv 0$, unless the same sources emit in different frequencies, as previously commented).
In this circumstance, redshift and astrometric measurements completely decouple, and the total SNR is simply the sum of the individual SNRs.

In the case of the noise covariance matrix, in this work we consider a purely diagonal noise matrix of the form~\cite{Caliskan:2023cqm}\footnote{
The redshift and angular displacement measurements appear to have a factor~$2$ difference.
This difference is easily explained.
Suppose to have measured a displacement~$\delta\mathbf{n}^\mathrm{meas} = \delta\mathbf{n} + \bm{\nu}$, where~$\bm{\nu}$ is the 2D vector of white noise.
Each of the two components of the noise is sampled from a Gaussian distribution~$\nu^i \sim \mathcal{N} (0,\sigma^\mathrm{tot}_\theta/\sqrt{2})$, in such a way that total error is~$\left\langle |\bm{\nu}|^2 \right\rangle = (\sigma^\mathrm{tot}_\theta)^2$.
Since in real space the noise components have expectation value
\begin{equation*}
    \left\langle \nu^i (\hat{n}) \nu^j (\hat{n}') \right\rangle = \frac{(\sigma^\mathrm{tot}_\theta)^2}{2} \delta^K_{ij} \delta^D(\hat{n}- \hat{n}'),
\end{equation*}
we have that in harmonic space the noise angular power spectrum is given by
\begin{equation*}
    \left\langle \nu^X_{\ell m}  \nu^Y_{\ell' m'} \right\rangle = \left\langle \int d\hat{n} d\hat{n}' \nu^i (\hat{n}) Y^{i,X}_{\ell m} (\hat{n}) \nu^j (\hat{n}') Y^{j,Y}_{\ell' m'} (\hat{n}') \right\rangle = \frac{(\sigma^\mathrm{tot}_\theta)^2}{2} \delta^K_{XY} \delta^K_{\ell\ell'} \delta^K_{mm'}.
\end{equation*}}
\begin{equation}
    \mathcal{N}_\ell = \mathrm{diag}\left( N^{zz}_\ell, N^{E_{(1,1)} E_{(1,1)}}_\ell, \cdots, N^{E_{(n,p)} E_{(n,p)}}_\ell, N^{B_{(1,1)} B_{(1,1)}}_\ell, \cdots, N^{B_{(n,p)} B_{(n,p)}}_\ell \right),
\end{equation}
where
\begin{equation}
    \begin{aligned}
        N^{zz}_\ell &= \frac{4\pi}{N_\mathrm{psr}} \int df |W_f(f,f_j,\Delta f_j)|^2 S_n(f), \\
        N^{X_{j,k}X_{j,k}} &= \frac{2\pi \sigma^2_\mathrm{tot}}{N_\mathrm{obj,j}},
    \end{aligned}
\end{equation}
and~$S_n(f)$ is the one-sided noise power spectrum, $N_\mathrm{obj,j}$ is the number of objects in the~$j$-th distance bin, and~$\sigma^2_\mathrm{tot}$ is the total error on the angular position, which is given by
\begin{align}
    \sigma_{\rm tot}^2 = \int_{1/T}^{1/\Delta t}df  \sigma_\theta^2 \Delta t\,.
\end{align}
In particular, for the astrometric noise, we assume a future-generation Gaia-like satellite for stellar observations and a next-generation Roman-like mission for asteroid measurements.
This choice is motivated by the extremely small magnitude of expected deflections and the need for a precise characterization of the velocities of the objects to robustly assess a detection.
In detail, we assume the same observation time and cadence, but we assume an improved angular sensitivity for each object, also starting from the estimates of refs.~\cite{Mentasti:2023gmr, Caliskan:2023cqm}:
\begin{align*}
    &{\rm GAIA-like}\quad & & \sigma_\theta^2= 3 \, \mu{\rm as}& & \Delta t =  14 \,{\rm days} & & T= 10\,{\rm years}\\
    &{\rm Roman-like}\quad & & \sigma_\theta^2= 0.1 \,\mu{\rm as} & &\Delta t = 30\,{\rm min}  & & T= 5\,{\rm months}
\end{align*}
In the case of pulsars, instead, we consider projected estimates for the SKAO sensitivity for 5 and 20 yrs of observation time~\cite{Shannon2026.SKA,SKAOPulsarScienceWorkingGroup:2025oyu}. 
Finally, to quantify the error future experiments will measure, we resort on a Fisher Matrix approach.
In particular, for each parameter of interest~$\theta_\alpha$, we have that the Fisher matrix reads as
\begin{equation}
    F_{\theta_\alpha\theta_\beta} = f_\mathrm{sky} \sum_\ell \frac{2\ell+1}{2} \mathrm{Tr}\left[ \left(\partial_{\theta_\alpha} \mathcal{C}_\ell \right) \left(\mathcal{C}_\ell+\mathcal{N}_\ell\right)^{-1} \left(\partial_{\theta_\beta} \mathcal{C}_\ell \right) \left(\mathcal{C}_\ell+\mathcal{N}_\ell\right)^{-1} \right],
\label{eq:FisherMatrix}
\end{equation}
where the matrix~$\partial_{\theta_\alpha} \mathcal{C}_\ell$ has for elements the derivatives of the angular power spectra with respect to the desired parameter.

Although astrometric datasets are well suited for searches of a stochastic GWB, observational limitations of both absolute and relative astrometric measurements still play a crucial role, as analyzed by Ref.~\citep{Vaglio:2025tex}. 
In the case of absolute astrometry, systematic errors are related to the satellite’s orientation: as the field of view rotates while the satellite moves along its orbit over the full observation period, correlations and uncertainties are introduced in the measured sky positions, potentially exceeding the precision achieved in a single observation within one field of view~\cite{2012A&A...543A..15H, Geyer:2024lvk}. 
Relative astrometry has the potential to mitigate this problem by considering relative angles between star pairs. 
However, a main limitation in this case arises from the small field of view of current satellites, e.g, Gaia. 
Recent works show that the change in the relative angle between astrophysical objects~$\psi$ induced by a GWB scales as~$\delta\psi\sim h\psi$; thus, it is suppressed for small fields of view~\cite{Crosta:2024udx, Vaglio:2025tex}. 
Nevertheless, as argued by Ref.~\citep{Vaglio:2025tex}, correlating astrometric and pulsar timing measurements remains a powerful and complementary way to detect GWs. 
For the purposes of this analysis, we neglect these systematic uncertainties in our forecasts; however, their effect could naturally be incorporated into the covariance matrix of both statistical estimators in equations~\eqref{eq:SNR2} and~\eqref{eq:FisherMatrix}. 
In both relative and absolute astrometry, accounting for their presence would degrade the angular sensitivity of each satellite and potentially increase the errors in the measurement of the GWB parameters.


\section{Results}
\label{sec:results}


\subsection{Full harmonic spectrum}

\begin{figure}[ht]
    \centerline{
    \includegraphics[width=1\linewidth]{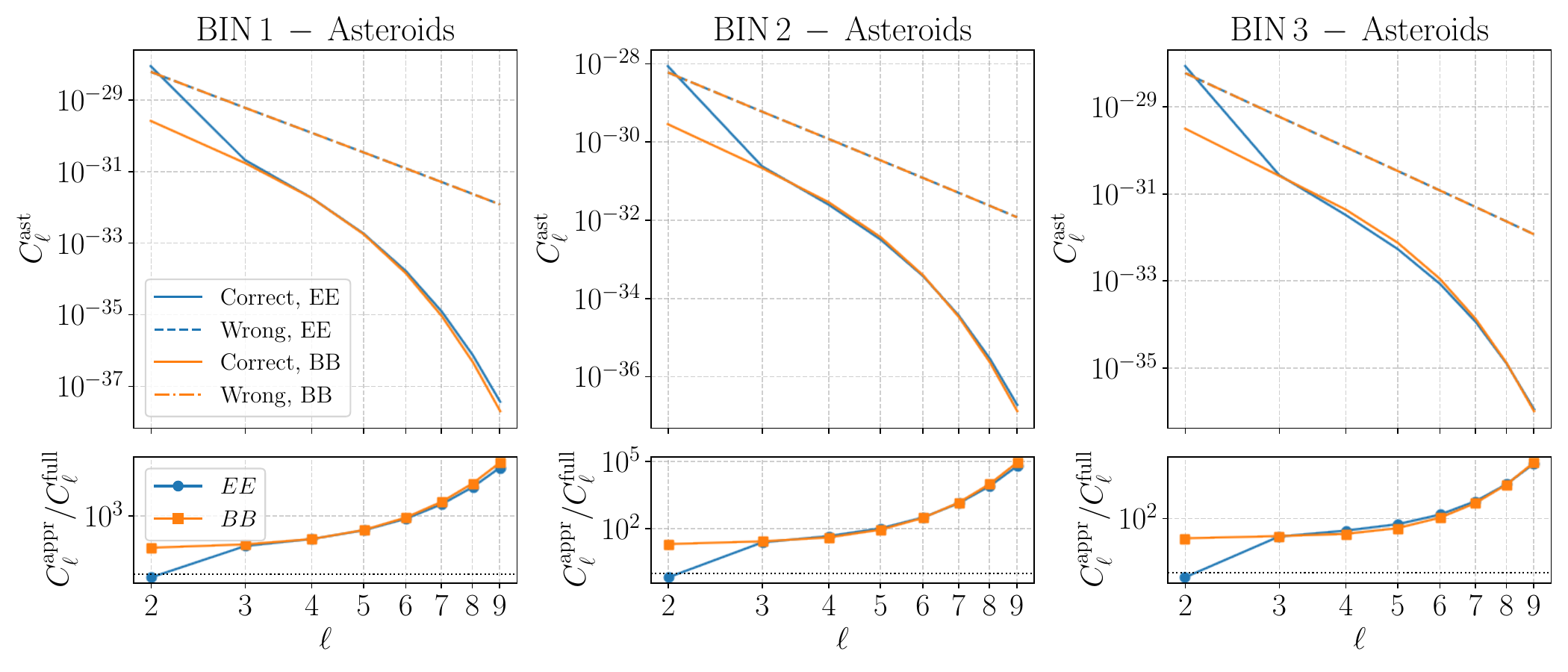}}
    \caption{\textit{Top panels:} angular power spectra for the asteroid distance bins considered in this work. 
    In each plot we compare the full approach that consider the finite distance effects (\textit{solid lines}) with the incorrect infinite-distance approximation (\textit{dashed lines}). 
    \textit{Bottom panels:} ratio of the incorrect-to-correct angular power spectra.
    The incorrect assumption of the infinite distance limit leads an overestimation of the signal by 3 to 5 orders of magnitude for the first multipoles.}
    \label{fig:asteroids_full}
\end{figure}

Although stars targeted by GAIA-like experiments are always in the infinite-distance regime, the same cannot be said for asteroids targeted by Roman, as also shown in the left panel of figure~\ref{fig::sources_dete}.
We compare in figure~\ref{fig:asteroids_full} the E- and B-mode angular power spectra obtained from equation~\eqref{Eq::Clrrp} with the angular power spectra computed in the infinite-distance limit for the population of main asteroids described in section~\ref{sec:observational_targets}.
As the figure shows, the discrepancy between the two approaches can be appreciated already starting from~$\ell=2$. 
Regarding the B-modes, we find that the long-distance approximation overestimates the power spectrum at any multipole; in particular, by 2 orders of magnitude at~$\ell=2$ and up to 5 orders of magnitude at~$\ell=9$. 
The E-modes, instead, show a different behavior: the infinite distance approximation underestimates the angular power spectrum with respect to the full approach at~$\ell=2$, but it overestimates the true value for~$\ell \geq 3$.
This behavior is consistent with the phenomenology of the E- and B-mode form factors showed in the lower panel of figure~\ref{fig:FEB2_of_x}.

This figure showcase the importance of our accurate modeling when dealing with nearby sources, as the asteroids in the Main Belt, whose spatial distribution spans a wide range of~$\tilde{x}\in [ 10^{-5} , 1]$. 
Observationally, the almost totality of the signal is comprised in the~$\ell=2$ multipoles, which is also the easiest to detect. 
Thus, in this case, underestimating the angular deflection by using the long-distance limit leads to a biased measure of both the mean value of and error on the GWB amplitude. 
In contrast, for higher multipoles, this mismodelling would lead to overestimate of many orders of magnitude of the spectrum. 
Interestingly, this behavior depends on the sources considered and frequencies probed. 
For instance, given a range of frequencies and a set of sources whose distance lies in the interval $\tilde{x}\in[1,50]$, the long-distance approach would underestimate the angular power spectra also for higher multipoles and not only for $\ell=2$, as can be understood from figure~\ref{fig:FEB2_of_x}.


\subsection{Signal-to-Noise results}

\begin{figure}[ht]
    \centering
    \includegraphics[width=0.6\linewidth]{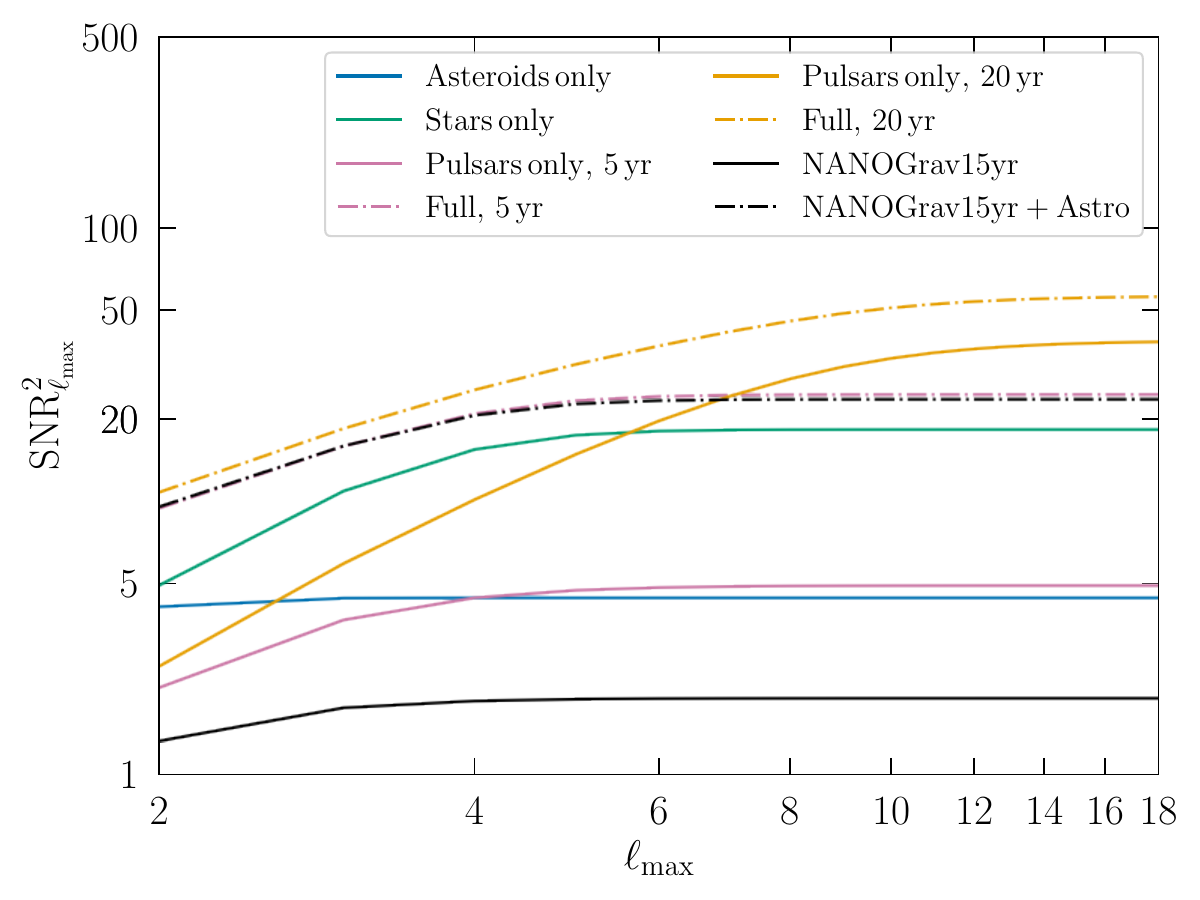}
    \caption{Cumulative angular SNR for different~$\ell_{\rm max}$. 
    The solid lines represent the cumulative SNR obtained considering stars, asteroids and redshift alone, while the dot-dashed line show the SNR obtained combining the probes and that results higher (as expected) than the single probes themselves. 
    When including pulsars we consider three benchmark cases: current PTA analysis with 67 pulsars and 15 years of observation time, and SKAO with 67 pulsars and 5 years of observation time and with 500 pulsars and 20 years of observation time.}    
    \label{fig:SNR_full}
\end{figure}

To understand the potential of cross-correlations between astrometric and redshift measurements, we evaluate the cumulative angular SNR for the different probes considered. 
For a given probe, we account for auto-correlation and cross-correlation spectra among the different bins and for the cross-correlations with different probes (e.g., stars' deflections with asteroids deflections). 
We also include the correlation with the redshift measurements, expected to give a valuable contribution. 
We inject the same GWB found by recent PTA analysis as an example and for each detector considered we adopt the noise specifications reported in section~\ref{sec::Statistical_Tools}. 
We report our findings in figure~\ref{fig:SNR_full} where we show the cumulative angular SNR accounting only for stars and asteroids. 
As shown in the figure, each probe taken individually is, in principle, capable of detecting the signal with a cumulative SNR that depends on the angular resolution of the instrument considered. 
In particular, our analysis indicates that a Roman-like detector with enhanced angular precision could achieve detections up to $\ell = 3$, and that a GAIA-like survey, monitoring the positions of distant stars, has encouraging detection prospects, with detectable contributions from multipoles up to~$\ell = 5$. 

The significance of the detection can be further improved in several ways. 
First, increasing the angular resolution enables a more accurate characterization of the signal and a more precise measurement of the deflection. 
Second, adopting a smaller observational cadence broadens the accessible frequency range, thereby extending the portion of the signal that can be probed and consequently enhancing the overall detectability. 
In addition, if also high multipoles are detectable, we would be able to further reduce the error on our GWB amplitude measurement. 
However, these experiments are constrained by a relatively large cadence, which reduces the accessible frequency range compared to Roman. 
Nevertheless, extending the mission total observation time would probe the low frequency regime, which necessarily requires dropping the long-distance approximation and adopting our result. 

In the case of pulsars, given the current findings and the timeliness of the results, we estimate the SNR with current pulsar timing data (NANOGrav with 15 years of observation time) and for the SKAO accounting for different numbers of pulsars and observation times.
We compute the SNR accounting only for pulsars and, separately, including all the probes and their correlations in the analysis.
Given the recent prospect, the SKAO telescopes is expected to monitor about~$200$ millisecond pulsars with a cadence of $\sim 15$ days; however, current estimates suggest the possible detection of up to 1000 millisecond pulsar for the design baseline, although not all of them might be suitable for a residua timing analysis. 
First, we find that also redshift measurements alone already yield a SNR larger than unity, as expected by the existing evidence of the Hellings-Downs correlation from current PTA datasets. However, with current PTA measurements, we observe that only the first multipoles are observed with a relatively low SNR. Given the improved sensitivity of SKAO, we find that already with the same number of pulsars currently monitored, after 5 yrs of observation time it will be possible not only to observe the signal, but also to detect multipoles up to~$\ell = 5$. 
In a more futuristic scenario, for an observation time of 20 years and 500 monitored pulsars, we find that SKA would sensibly improve the detection significance of the GWB by being able to observe up to~$\ell\simeq 13$. 

Finally, we perform a complete analysis including all the probes and their cross-correlations. 
We report the corresponding lines (dot-dashed) in the figure, considering the two different benchmark cases for pulsars observations. 
In all cases, we find that including all possible cross 
-correlations among the different probes enhances the angular SNR. 
The most significant effect appears at the lowest multipoles: the SNR is substantially increased compared to the single-probe cases and in the case of current PTA and SKAO with 5 year of observation time is almost comparable.
For the first SKAO benchmark scenario (67 pulsars over 5 years), and under the noise assumptions of a future GAIA-like detector, we predict sensitivity up to $\ell = 5$.  
For pulsars alone, the dominant contributions arise from $\ell = 2$ and $\ell = 3$ (with a minor contribution from $\ell = 4$), whereas for asteroids the primary contribution to the SNR comes exclusively from $\ell = 2$. 
When combining all probes, the power in the lowest multipoles increases relative to the individual-probe cases up to $\ell = 4$. 
At higher multipoles, where the signal is dominated by star deflections measurements (given the noise specifications adopted in this analysis), the SNR follows the same trend as the stellar-deflection signal, with an offset quantifying the gain obtained from combining all probes.

On the other hand, for the second benchmark scenario (500 pulsars over 20 years), we find a similar behavior. 
However, given the improved sensitivity of the SKAO in this case pulsars observations dominate the signal at higher multipoles. 
Nevertheless, we find a substantial increase up to $\ell =5$ that comes from the correlation of different probes and highlights the need to take a synergistic approach when looking at the information from the GW probes. 


\subsection{Fisher forecast results}

Finally, we perform a Fisher forecast injecting the three GW signals discussed in section~\ref{subsec::GW_models}, in order to assess the potential of a synergistic effort between astrometric and pulsar timing experiments to detect a GWB. 
In the case of a powerlaw-like spectrum as the one detected by current PTA data or as the one produced by supermassive black hole binaries, we include in our forecast the amplitude and the spectral index. 
Instead, in the case of a flat spectrum we include only the amplitude in the analysis. 

\begin{figure}[ht]
    \centering

    \begin{minipage}{0.48\linewidth}
        \centering
        \includegraphics[width=1\linewidth]{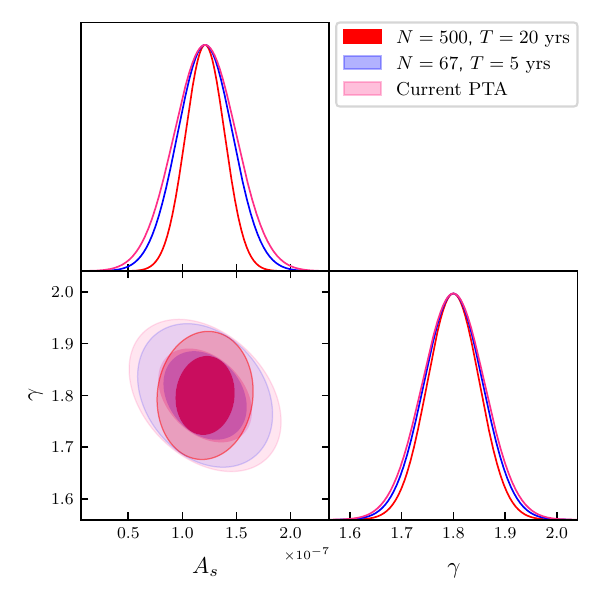}
        \label{fig:fisher_pta}
    \end{minipage}
    \hfill
    \begin{minipage}{0.48\linewidth}
        \centering
        \includegraphics[width=1\linewidth]{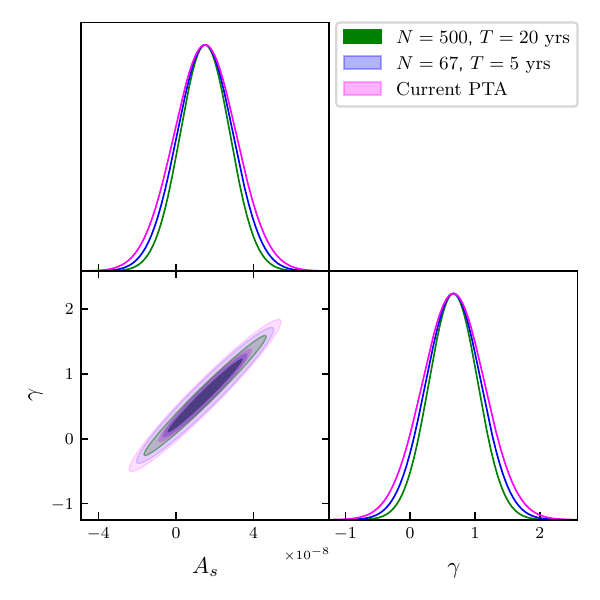}
        \label{fig:fisher_agwb}
    \end{minipage}

    \caption{Corner plots showing the error on the amplitude and spectral index of the benchmark GWB reported in Sec.~\ref{subsec::GW_models}, obtained with a Fisher analysis.}
    \label{fig:fisher_combined}
\end{figure}

Our findings for the first two benchmark cases are reported in figure~\ref{fig:fisher_combined}. 
Since we have already emphasized the importance of combining the different probes in the SNR analysis, we perform the forecast including directly all the auto- and cross-correlations for each probe as explained in section~\ref{sec::Statistical_Tools}.
However, we emphasize that the inclusion of astrometric measurements plays a fundamental role in the result we report below. This effect can be understood by considering~\eqref{eq::Cl_zz} with the power-law GW signal considered in this analysis. It is straightforward to show that considering only PTA measurements the entries of the Fisher matrix would be $\frac{dC_\ell}{dA}\sim \frac{C_\ell}{A} \sim \frac{(l-2)!}{(l+2)!} \int   \frac{{\rm d}f}{f^3} \left(\frac{f}{f_s}\right)^\gamma$ and $\frac{dC_\ell}{d\gamma}\sim \frac{(l-2)!}{(l+2)!} \int \frac{{\rm d}f}{f^3} A\left(\frac{f}{f_S}\right)^\gamma \log\left(\frac{f}{f_s}\right)$. The mild logarithmic dependence in the latter derivative would not be fully appreciated if the frequency range considered is too short. Thus the entries of the Fisher matrix would be almost similar, leading to an almost degenerate matrix. This degeneracy is broken with the inclusion of astrometric measurements (which probe different frequency bands) leading to the results we report below.

We repeated the analysis for three different benchmark cases, first with current PTA observations (67 pulsar and 15 years of observation time), then with SKAO with 67 pulsars and for 5 years of observation time and considering 500 pulsars for 20 years of observation time. We observe improved constraints when considering SKAO benchmark cases compared to current PTA measurements. The best performance is achieved with SKA data assuming 20 years of observations and 500 monitored pulsars. We find that errors obtained with current PTA measurements are comparable, albeit slightly larger, than those obtained with SKAO already after 5 years of observations.
For the generic PTA-like signal, we find that in all cases it is possible to detect the amplitude with an error of less than the 50\% of the central value. 
When the number of pulsars and observation time are increased, the error decreases to less than 40\%, due to the increased sensitivity of pulsar timing.
In detail, we find $\sigma_A=0.3\times10^{-7}$ with current PTA measurements, $\sigma_A=0.25\times10^{-7}$ in the first SKAO benchmark case and $\sigma_A=0.18\times10^{-7}$ in the latter. 
Regarding the spectral index, instead, our results suggest that future astrometric and redshift measurements would constrain it with an error of the percent, respectively $\sigma_\gamma/\gamma=0.056/1.8\simeq3\%$ and $\sigma_\gamma/\gamma=0.05/1.8\simeq2.7 \%$. When considering current PTA measurements, instead, we find $\sigma_\gamma/\gamma=0.06/1.8\simeq3.3\%$.
The enhanced constraining power on the spectral index, as opposed to the amplitude, can be understood as follows: when the amplitude is fixed at a reference frequency, the spectral index determines the relative weighting of power across the entire relevant frequency band. 
Consequently, even a small change in the spectral index produces appreciable modifications in the spectrum, correspondingly leading to a stronger impact on the analysis, as will also be explained in the following.

In the case of a GWB generated by super-massive black hole binaries, we find slightly worse results. 
This is due to the lower amplitude of the GW considered (with a decrease of about one order of magnitude, given the latest PTA constraints). 
We find $\sigma_A=1.6 \times10^{-8}$ and $\sigma_\gamma = 0.479$ with current PTA measurements and $\sigma_A=1.4 (1.3) \times10^{-8}$ and $\sigma_\gamma = 0.427 (0.378)$ in the two SKAO cases considered. 
Finally, in the case of a flat (cosmological) background, we forecast a relative error of the 20\%, 18\% and 13\% respectively for the same cases considered before, even though the amplitude is now lower than in the first case and comparable to the second. 
Specifically we find $\sigma_A=2.1\times10^{-9}$ when considering current PTA measurements and $\sigma_A=1.88\times10^{-9}$ and $\sigma_A=1.3\times10^{-9}$ for the two SKAO benchmark cases considered. 
This improvement can be understood by noting that, for a flat spectrum, all frequencies contribute equally to the signal and in addition, having one less parameter reduces the degeneracies.
Finally, we remark  that in all of the cases considered SKAO will outperform current PTA analysis already with an observation time of 5 years and the same number of pulsars currently monitored.


\section{Conclusions}
\label{sec:conclusions}

In this work we studied in detail the potential of astrometric surveys and their synergy with pulsar timing arrays. 
First, we presented a new approach which allows to evaluate the astrometric deflection for any frequency, without limitations on the distance of the source considered, i.e., for any value of~$\tilde{x}=2\pi f r$.
This result represents the main theoretical novelty of this work.
We also introduced the form factors for the E- and B-mode in equation~\eqref{Eq:ClBB_ClEE}, which asymptotically reach~$(-1)^\ell$ in the infinite distance limit~$\tilde{x}\to \infty$ and show a non-trivial dependence for finite value of~$\tilde{x}$ otherwise. 
Our results show that for the E-mode when~$\ell=2$ the form factor reaches a higher value with respect to the long-distance approximation, highlighting the need of a full approach. 
Including the dependence on distance enables the modeling of the three-dimensional distribution of the targeted sources, allowing for a more comprehensive estimate of the deflection induced by a GWB. 
We also demonstrated that a fully consistent treatment can lead to either an overestimation or underestimation of the harmonic power spectrum. 
Additionally, it allows to evaluate all possible correlations between different observational probes, including redshift measurements obtained from pulsar timing arrays. 
Furthermore we show how the distance–frequency plane is sampled by current and future surveys using both astrometric and redshift data and we provide a simple mapping, given a survey, to understand the frequency range spanned and the sources to be considered corresponding to different values of $\tilde{x}$. 

On the astrometric side, in this work we considered the distribution of asteroids in the Main Belt and stars in the Milky Way, assuming a GAIA-like and Roman-like detector to monitor their position. 
For the redshift measurements, instead, we consider 67 pulsars monitored for 15 years accordingly to current PTA measurements, 67 pulsars monitored for 5 years by SKA for 5 years and 500 pulsars monitored for 20 years.
We performed an SNR analysis first only considering the single probes and then fully including their correlation, finding in all cases a SNR greater that one showing that future observation the astrometric deflection would be fundamental for the characterization of a GWB. 
We specify that our results are dependent on the angular sensitivity considered for the different satellites considered and on the number of sources monitored. 
Furthermore, we estimate the capability of this method to constrain a GWB with a Fisher analysis. 
Considering the latest results from PTA, we consider as benchmark cases the GWB responsible for the Hellings-Downs pattern observed, the upper bounds on the background given by SMBHB in their inspiralling phase and a flat GWB. 
Depending on the spectral index and on the amplitude, we find a percentage error going from less than the $50\%$ of the central value up to the $2.7\%$ in the optimal scenario.

Although in this work we focus exclusively on asteroids populating the Main Belt, other asteroid families are also present in the Solar System and could potentially be monitored to detect the presence of a background.
However, their current known number of is at least one/two orders of magnitude smaller than that of Main Belt asteroids, also for the Kuiper Belt asteroids, where we found only~$N_\mathrm{ast} \approx 4100$ with known orbital motion\footnote{The total Small-Body Database contains approximately~$1.47\times 10^6$ objects with known orbital motion, i.e., asteroids not in the Main Belt represent only~$10\%$ of the database.}.
Some of these limitations are purely observational, and future missions would very likely increase these numbers, since it might be the case that there are around~$10^5$ objects in that region.

In addition, we note that our analysis assumes an ideal reconstruction of the angular displacement, neglecting the influence of the sources' peculiar velocities or orbital motion.
This constitutes a primary limitation of the method, whose detailed modeling is beyond the scope of the present work and we leave to a future more detailed analysis.


\acknowledgments

The authors thank Pau Ramos for useful discussions.
NB acknowledges support by PRD/ARPE 2022 ``Cosmology with Gravitational waves and Large Scale Structure - CosmoGraLSS''.
GP \& DB acknowledge support from the COSMOS network (www.cosmosnet.it) through ASI (Italian Space Agency) Grants 2016-24-H.0, 2016-24-H.1-2018 and 2020-9-HH.0


\appendix 
\section{Mathematical notation and conventions}
\label{app:notation_conventions}


\subsection{Fourier transform}

In this work, we adopt the following convention for the Fourier transform
\begin{equation}
    f(\mathbf{x}) = \int \frac{d^3k}{(2\pi)^3}  f(\mathbf{k}) e^{i\mathbf{k}\cdot\mathbf{x}} , \qquad f(\mathbf{k}) = \int d^3x f(\mathbf{x}) e^{-i\mathbf{k}\cdot\mathbf{x}}.
\end{equation}
In the case of a real scalar field, for which~$f^*(\mathbf{k})=f(-\mathbf{k})$, the two-point function in Fourier space reads as
\begin{equation}
    \left\langle f(\mathbf{k}_1) f^*(\mathbf{k}_2) \right\rangle = (2\pi)^3 \delta^D(\mathbf{k}_1-\mathbf{k}_2) P(k),
\end{equation}
where~$P(k)$ is the power spectrum.
Therefore, in the case of real tensor field, as for the case of a GW perturbation moving at the speed of light, we have
\begin{equation}
    \begin{aligned}
        h_{ij}(t,\mathbf{x}) &= \int \frac{d^3k}{(2\pi)^3} h_{ij}(k,\mathbf{k}) e^{-ik(t - \hat{\mathbf{k}}\cdot\mathbf{x})} = \sum_\lambda \int \frac{d^3k}{(2\pi)^3} h_\lambda(k,\mathbf{k}) e_{ij}^\lambda(\hat{\mathbf{k}}) e^{-ik(t - \hat{\mathbf{k}}\cdot\mathbf{x})} \\
        &= \sum_\lambda \int df d\hat{\mathbf{p}} \ h_\lambda(f,\hat{\mathbf{p}}) e_{ij}^\lambda(\hat{\mathbf{p}}) e^{- 2\pi i f (t - \hat{\mathbf{p}} \cdot \mathbf{x})},
    \end{aligned}
\end{equation}
after the change of variable~$k_j=2\pi f_j$ such that~$d^3k = (2\pi)^3 f^2 df d\hat{\mathbf{p}}$, and a redefinition of~$h_\lambda$.


\subsection{Scalar and vector spherical harmonics}

In this work, the decomposition in spherical harmonics of a scalar function follows the convention
\begin{equation}
    X(\hat{\mathbf{n}}) = \sum_{\ell m} X_{\ell m} Y_{\ell m} (\hat{\mathbf{n}}), \qquad X_{\ell m} = \int d\hat{\mathbf{n}} X(\hat{\mathbf{n}}) Y^*_{\ell m}(\hat{\mathbf{n}}),
\end{equation}
where~$X_{\ell m}$ are the harmonic coefficients, the spherical harmonics are explicitly defined as
\begin{equation}
    Y_{\ell m}(\hat{\mathbf{n}}) = \sqrt{\frac{2\ell+1}{4\pi} \frac{(\ell-m)!}{(\ell+m)!}} P_\ell^m(\cos\theta) e^{im\phi},
\end{equation}
and~$P_\ell^m(\cos\theta)$ are the associated Legendre polynomials.
This definition ensures that spherical harmonics are an orthonormal basis, i.e.,
\begin{equation}
    \int d\hat{\mathbf{n}} Y_{\ell m}(\hat{\mathbf{n}}) Y_{\ell' m'}(\hat{\mathbf{n}}) = \delta^K_{\ell\ell'} \delta^K_{mm'}.
\end{equation}

On the other hand, a 2D vector tangential to a sphere of radius~$r$ can be decomposed into its ``electric'' and ``magnetic'' components, also known as E- and B-modes, as
\begin{equation}
    \mathbf{X}(\hat{\mathbf{n}}) = \sum_{\ell m} \left[ E_{\ell m} \mathbf{Y}^{E}_{\ell m} (\hat{\mathbf{n}}) + B_{\ell m} \mathbf{Y}^{B}_{\ell m} (\hat{\mathbf{n}}) \right],
\end{equation}
where the vector spherical harmonics are defined as~\cite{Varshalovich:1988ifq}
\begin{equation}
    \begin{aligned}
        \mathbf{Y}^E_{\ell m}(\hat{\mathbf{n}}) &= \frac{1}{\sqrt{\ell(\ell+1)}} \nabla_\Omega Y_{\ell m}(\hat{\mathbf{n}}) = \frac{1}{\sqrt{\ell(\ell+1)}} r\nabla Y_{\ell m}(\hat{\mathbf{n}}), \\
        \mathbf{Y}^B_{\ell m}(\hat{\mathbf{n}}) &= \frac{-i}{\sqrt{\ell(\ell+1)}} (\hat{\mathbf{n}} \times \nabla_\Omega) Y_{\ell m}(\hat{\mathbf{n}}) = \frac{-i}{\sqrt{\ell(\ell+1)}} (\hat{\mathbf{n}} \times r\nabla) Y_{\ell m}(\hat{\mathbf{n}}), 
    \end{aligned}
\end{equation}
and~$\nabla_\Omega$ is the angular part of the~$\nabla$ operator.
The inverse transformation for electric and magnetic components is given by
\begin{equation}
    E_{\ell m} = \int d\hat{\mathbf{n}}\ \mathbf{X} (\hat{\mathbf{n}}) \cdot \mathbf{Y}^{E*}_{\ell m} (\hat{\mathbf{n}}), \qquad B_{\ell m} = \int d\hat{\mathbf{n}}\ \mathbf{X} (\hat{\mathbf{n}}) \cdot \mathbf{Y}^{B*}_{\ell m} (\hat{\mathbf{n}}). 
\end{equation}
Also in this case, vector spherical harmonics are an orthonormal basis, for which
\begin{equation}
    \int d\hat{\mathbf{n}} \mathbf{Y}^I_{\ell m}(\hat{\mathbf{n}}) \cdot \mathbf{Y}^J_{\ell' m'}(\hat{\mathbf{n}}) = \delta^K_{IJ} \delta^K_{\ell\ell'} \delta^K_{mm'}.
\end{equation}

Since by construction the vector~$\mathbf{X}$ is tangential to the sphere, i.e., $\mathbf{X} \cdot \hat{\mathbf{n}} = \mathbf{X} \cdot \nabla r = 0$, the computation of the electric and magnetic components reduces to
\begin{equation}
    \begin{aligned}
        E_{\ell m} &= \int d\hat{\mathbf{n}}\ \mathbf{X} \cdot \mathbf{Y}^{E*}_{\ell m} = \frac{1}{\sqrt{\ell(\ell+1)}} \int d\hat{\mathbf{n}} \left [ \nabla \cdot (\mathbf{X}\ r Y^*_{\ell m}) - (\mathbf{X} \cdot \nabla r) Y^*_{\ell m} - (r\nabla \cdot \mathbf{X}) Y^*_{\ell m} \right] \\
        &= -\frac{1}{\sqrt{\ell(\ell+1)}} \int d\hat{\mathbf{n}} Y^*_{\ell m}(\hat{\mathbf{n}}) (r\nabla \cdot \mathbf{X}), \\
        B_{\ell m} &= \int d\hat{\mathbf{n}} \ \mathbf{X} \cdot \mathbf{Y}^{B*}_{\ell m} = \frac{i}{\sqrt{\ell(\ell+1)}} \int d\hat{\mathbf{n}} \left[ \nabla \cdot (\mathbf{X}\times \hat{\mathbf{n}}\ r Y^*_{\ell m}) - rY^*_{\ell m} \hat{\mathbf{n}} \cdot \nabla \times \mathbf{X} \right. \\
        &\qquad\qquad\qquad\qquad\qquad\qquad\qquad\qquad\qquad \left. + rY^*_{\ell m} \mathbf{X} \cdot \nabla \times \hat{\mathbf{n}} + Y^*_{\ell m} \mathbf{X} \cdot \nabla r \times \hat{\mathbf{n}} \right] \\
        &= \frac{-i}{\sqrt{\ell(\ell+1)}} \int d\hat{\mathbf{n}} Y^*_{\ell m} (\hat{\mathbf{n}})\ (\hat{\mathbf{n}} \times r\nabla) \cdot \mathbf{X},
    \end{aligned}
\end{equation}
by using the periodicity of the spherical harmonics functions to cancel the surface contributions, and the identities~$\nabla \times \hat{\mathbf{n}}=0$, $\nabla r \times \hat{\mathbf{n}} = 0$ and~$\mathbf{A} \cdot (\nabla \times \mathbf{B}) = (\mathbf{A}\times\nabla)\cdot \mathbf{B}$ to further simplify the magnetic component integral.


\subsection{Legendre polynomials}

In this work, we make extensive use of the Legendre ($P_\ell$) and associated Legendre ($P^m_\ell$) polynomials and their properties, which we summarize here for later convenience.
The two classes of polynomials are connected by the relation
\begin{equation}
    P^m_\ell(\mu) = (-1)^m (1-\mu^2)^{m/2} \frac{d^m P_\ell(\mu)}{d\mu^m},
\end{equation}
and follow
\begin{equation}
    P_{\ell}^{-m}(\mu) = (-1)^m \frac{(\ell-m)!}{(\ell+m)!} P_\ell^m(\mu)\,, \quad
    P_\ell^m (1) = \delta^K_{m,0}\,, \quad
    P_\ell^m(-\mu) = (-1)^\ell P_\ell^m(\mu) \,.
\end{equation}
However, in the following, we are mostly interested in the~$m=2$ case, for which
\begin{equation}
    P^2_\ell(\mu) = (1-\mu^2) \frac{d^2 P_\ell(\mu)}{d\mu^2}, \qquad P^{-2}_\ell(\mu) = \frac{(\ell-2)!}{(\ell+2)!} P^2_\ell(\mu).
\end{equation}

Legendre polynomials are the solution of the differential equation
\begin{equation}
    (1-\mu^2) \frac{d^2 P_\ell(\mu)}{d\mu^2} - 2\mu \frac{dP_\ell(\mu)}{d\mu} + \ell(\ell+1) P_\ell(\mu) = 0\,,
\end{equation}
and also form an orthogonal basis such that
\begin{equation}
    \int d\mu P_\ell(\mu) P_{\ell'}(\mu) = \frac{2}{2\ell+1} \delta^K_{\ell\ell'}.
\end{equation}
Additionally, they also satisfy the following relations
\begin{equation}
    \begin{aligned}
        \frac{\mu^2-1}{\ell} \frac{dP_\ell(\mu)}{d\mu} &= \mu P_\ell(\mu)-P_{\ell-1}(\mu), \\
        \frac{dP_{\ell-1}(\mu)}{d\mu} &= (\ell+1)P_\ell(\mu) + \mu \frac{d P_\ell(\mu)}{d\mu}, \\
        (\ell + 1) P_{\ell+1}(\mu) &= (2\ell+1) \mu P_\ell(\mu) - \ell P_{\ell-1}(\mu). \\
    \end{aligned}
\end{equation}
Finally, at the boundary of their domain of definition, Legendre polynomials and their first derivatives take the values
\begin{equation}
    P_\ell(1)=1, \quad P_\ell(-1)=(-1)^\ell, \quad
    \left. \frac{dP_\ell}{d\mu} \right|_{\mu=1} = \frac{\ell(\ell+1)}{2}, \quad \left. \frac{dP_\ell}{d\mu} \right|_{\mu=-1} = (-1)^{\ell+1} \frac{\ell(\ell+1)}{2},
\end{equation}
which can be easily derived using their symmetry property~$P_\ell(-y) = (-1)^\ell P_\ell(y)$.

For the purpose of this paper, it is useful to introduce two classes of functions,~$J_n$ and~$K_n$, defined as
\begin{equation}
    \begin{aligned}
        J_n &= \int d\mu\ \mu^n \frac{d^2P_\ell}{d\mu^2} = \left[\frac{\ell(\ell+1)}{2} - n\right] \left[1 + (-1)^{\ell+n} \right] + n (n-1) \int d\mu\ \mu^{n-2} P_\ell(\mu), \\
        K_n(w) &= \int d\mu\ \mu^n e^{iw\mu} \frac{d^2P_\ell}{d\mu^2} =  (-i)^n \frac{d^n K_0}{dw^n}, \\
    \end{aligned}
\end{equation}
where all the terms proportional to~$\delta^K_{\ell 0}$ and~$\delta^K_{\ell 1}$ are not reported, and~$K_0$ is
\begin{equation}
    \begin{aligned}
        K_0(w) &= \int d\mu\ e^{iw\mu} \frac{d^2 P_\ell}{d\mu^2} \\
        &= \frac{\ell(\ell+1)}{2} \left[ e^{iw} + (-1)^\ell e^{-iw} \right] - i w \left[ e^{iw} + (-1)^{\ell+1} e^{-iw} \right] -2 i^\ell w^2 j_\ell(w),
    \end{aligned}
\end{equation}
obtained by implementing the Legendre decomposition of the complex exponential
\begin{equation}
    \int d\mu\ e^{iw\mu} P_\ell(\mu) = \int d\mu \left[ \sum_{\ell'} i^{\ell'} (2\ell'+1) j_{\ell'}(w) P_{\ell'}(\mu) \right] P_\ell(\mu) = 2 i^\ell j_\ell(w), \\
\end{equation}
where~$j_\ell(w)$ is the spherical Bessel function.


\subsection{Notable identities}

First, we report some generic result that comes in hand when computing the harmonic coefficients of the angular displacement.
In particular, we make extensive use of the fact that
\begin{equation}
    \nabla_i r = n^i, \quad r\nabla_i n^j = \nabla_{\Omega,i} n^j = \delta^K_{ij} - n_i n_j, \quad r\nabla_i (\hat{\mathbf{p}} \cdot \hat{\mathbf{n}}) = p^i - (\hat{\mathbf{p}} \cdot \hat{\mathbf{n}}) n^i.
\end{equation}
Using these relations, it is straightforward to see that
\begin{equation}
    r\nabla_i \left[ 2\pi fr (1+ \hat{\mathbf{p}}\cdot\hat{\mathbf{n}}) \right] = 2\pi fr (n^i + p^i),
\end{equation}
since~$\partial r/\partial x^i = x^i/r = n^i$ and~$\partial (\hat{\mathbf{p}} \cdot \mathbf{r})/ \partial x^i = p^i$.
Therefore, by introducing the variable~$\Gamma = 2\pi fr (1+ \hat{\mathbf{p}}\cdot\hat{\mathbf{n}})$, we have that the spatial derivative of the distance form factor reads as
\begin{equation}
    \begin{aligned}
        r\nabla_i \mathcal{D} &= \partial_\Gamma \mathcal{D}\times r\nabla_i \Gamma = \frac{i\Gamma e^{i\Gamma} - e^{i\Gamma}+1}{\Gamma^2} 2\pi fr (n^i + p^i).
    \end{aligned}
\end{equation}

Some common integrals over the polar angle appearing in the calculations presented in appendix~\ref{app:explicit_Cl_calculation} are
\begin{equation}
    \begin{aligned}
        \int d\cos\theta (1-\cos\theta) P^2_\ell(\cos\theta) &= \int d\mu (1-\mu-\mu^2+\mu^3) \frac{d^2P_\ell(\mu)}{d\mu^2} \\
        &= J_0-J_1-J_2+J_3 = 4(-1)^\ell        
    \end{aligned}
\end{equation}
On the other hand, we have that some important relations involving the azimuthal angle are
\begin{equation}
\label{eq::int_phi}
    \int d\phi \cos(2\phi) e^{-im\phi} = \pi (\delta^K_{m,2} + \delta^K_{m,-2}), \qquad \int d\phi \sin(2\phi) e^{-im\phi} = -i\pi (\delta^K_{m,2} - \delta^K_{m,-2}),
\end{equation}
Finally, regarding the summation properties of Kronecker deltas, we have that
\begin{equation}
\label{eq::int_phi2}
    \sum_m \left(\delta^K_{m,2} + \delta^K_{m,-2} \right)^2 = \sum_m \left( \delta^K_{m,2} - \delta^K_{m,-2} \right)^2 = 2,\qquad \sum_m \left(\delta^K_{m,2} + \delta^K_{m,-2}\right)\left(\delta^K_{m,2} - \delta^K_{m,-2}\right) = 0.
\end{equation}


\section{Derivation of the angular power spectra}
\label{app:explicit_Cl_calculation}


\subsection{Selection of a reference frame}

Although equations~\eqref{eq:observed_angular_displacement} and~\eqref{eq:redshift_timing_residuals} are valid for any polarization degree of freedom, here we concentrate only on the General Relativity case, where~$\lambda=\{ +,\times \}$.
Therefore, since the GW propagation vector is orthogonal to the polarization tensor, and the latter is traceless, we have that
\begin{equation}
    p^je^{+,\times}_{ij} = 0, \qquad \mathrm{Tr}\ e^{+,\times} = 0.
\end{equation}
First, we conveniently choose a reference frame where the GW propagation happens along one of the axis and the line-of-sight is generic, so that
\begin{equation}
    \hat{\mathbf{p}} = (0,0,1), \qquad \hat{\mathbf{n}} = (\sin\theta\cos\phi, \sin\theta\sin\phi, \cos\theta).
\label{eq:frame_selection}
\end{equation}
In this frame, the polarization tensors read as
\begin{equation}
    e^+ = \begin{pmatrix}
    1 & 0 & 0 \\
    0 & -1 & 0 \\
    0 & 0 & 0 \\
    \end{pmatrix}, \qquad 
    e^\times = \begin{pmatrix}
    0 & 1 & 0 \\
    1 & 0 & 0 \\
    0 & 0 & 0 \\
    \end{pmatrix} ,
\end{equation}
therefore, the explicit form of the ``antenna patterns'' typical of the observables analyzed in this work are
\begin{equation}
\label{eq::tens_proj}
    \begin{aligned}
        \frac{n^j n^k e^+_{jk}}{1+\hat{\mathbf{p}}\cdot\hat{\mathbf{n}}} &= (1-\cos\theta) \cos(2\phi) , \qquad
        \frac{n^j n^k e^\times_{jk}}{1+\hat{\mathbf{p}}\cdot\hat{\mathbf{n}}} = (1-\cos\theta) \sin(2\phi), \\
        \frac{\varepsilon_{ijk} p^i n^j n^q e^+_{qk}}{1+\hat{\mathbf{p}}\cdot\hat{\mathbf{n}}} &= (1-\cos\theta) \sin(-2\phi) , \qquad
        \frac{\varepsilon_{ijk} p^i n^j n^q e^\times_{qk}}{1+\hat{\mathbf{p}}\cdot\hat{\mathbf{n}}} = (1-\cos\theta) \cos(-2\phi).
    \end{aligned}
\end{equation}

The choice made in equation~\eqref{eq:frame_selection} appears to be very specific, and certainly influences the value of the inferred harmonic coefficients.
Therefore, one might wonder whether the value of the two-point function depends on the choice of reference frame. 
In the case, the answer is negative, their value does not depend on any explicit choice of reference frame, as long as the observable~$X(\hat{\mathbf{n}})$ is invariant under rotations, i.e., $X'(\hat{\mathbf{n}}')=X(\hat{\mathbf{n}})$.
To prove this fact, let us consider how the harmonic coefficients transform under rotations
\begin{equation}
    X'_{\ell m} = \int d\hat{\mathbf{n}}' X'(\hat{\mathbf{n}}') Y^*_{\ell m} (\hat{\mathbf{n}}') = \sum_{n=-\ell}^{\ell} \left(D^\ell_{mn}\right)^* \int d\hat{\mathbf{n}} X(\hat{\mathbf{n}}) Y^*_{\ell n} (\hat{\mathbf{n}}) = \sum_{n=-\ell}^{\ell} \left(D^\ell_{mn}\right)^* X_{\ell n},
\end{equation}
where~$D^\ell_{mn}$ are the Wigner D-functions.
Therefore, in terms of the angular power spectrum, we have
\begin{equation}
    \begin{aligned}
        \left\langle X'_{\ell m} \left(X'_{\ell' m'} \right)^* \right\rangle &= \delta^K_{\ell\ell'} \delta^K_{mm'} C'_\ell \\
        &= \sum_{nn'} \left(D^\ell_{mn}\right)^* D^{\ell'}_{m'n'} \left\langle X_{\ell n} \left(X_{\ell' n'} \right)^* \right\rangle
        = \sum_{nn'} \left(D^\ell_{mn}\right)^* D^{\ell'}_{m'n'} \delta^K_{\ell\ell'} \delta^K_{nn'} C_\ell \\
        &= \sum_{n} \left(D^\ell_{mn}\right)^* D^{\ell}_{m'n} \delta^K_{\ell\ell'} C_\ell = \delta^K_{\ell\ell'} \delta^K_{mm'} C_\ell,
    \end{aligned}
\end{equation}
using the properties of the Wigner D-functions, i.e., the angular power spectra are invariant under rotations.
In practice, in this paper all the observable~$X(\hat{\mathbf{n}})=z, r\nabla\cdot\delta\mathbf{n}, (\hat{\mathbf{n}} \times r\nabla) \cdot \delta\mathbf{n}$ are scalar.
Aside from the redshift, which is evidently a scalar, for the angular displacement we observe that under a rotation represented by an orthogonal matrix~$R$ we have 
\begin{equation}
    \begin{aligned}
        r\nabla'\cdot\delta\mathbf{n}' &= (r\nabla')^T\delta\mathbf{n}' = [R(r\nabla)]^T (R\delta\mathbf{n}) = (r\nabla)^T (R^T R) \delta\mathbf{n} = r\nabla \cdot\delta\mathbf{n}, \\
        (\hat{\mathbf{n}}' \times r\nabla') \cdot \delta\mathbf{n}' &= \left[R\hat{\mathbf{n}} \times R(r\nabla)\right]^T (R\delta\mathbf{n}) = \left[R\left(\hat{\mathbf{n}} \times r\nabla\right)\right]^T (R\delta\mathbf{n}) \\
        &= \left[\hat{\mathbf{n}} \times r\nabla\right]^T (R^T R) \delta\mathbf{n} = (\hat{\mathbf{n}} \times r\nabla) \cdot \delta\mathbf{n}.
    \end{aligned}
\end{equation}
Alternatively, we can directly demonstrate the invariance of the redshift and angular displacement functions: for instance, in the case of the redshift, we see that the antenna patterns are such that
\begin{equation}
    F'_\lambda(\hat{\mathbf{n}}') = \frac{(\hat{n}')^T (e^\lambda)' \hat{n}'}{1+(\hat{p}')^T\hat{n}'} = \frac{(R\hat{n})^T (R e^\lambda R^T) (R\hat{n})}{1+(R\hat{p})^T(R\hat{n})} = \frac{\hat{n}^T e^\lambda \hat{n}}{1+\hat{p}^T\hat{n}} = F_\lambda(\hat{\mathbf{n}}).
\end{equation}
Similar relations can also be explicitly checked for the angular displacement, proving that our choice of reference frame is not affecting the estimate of the angular power spectra.


\subsection{Angular displacement angular power spectrum}

Applying the general definitions of appendix~\ref{app:notation_conventions}, we have that
\begin{equation}
\label{eq::E_B_mode}
    \begin{aligned}
        E^{f,r}_{\ell m} &= -\frac{1}{\sqrt{\ell(\ell+1)}} \int d\hat{\mathbf{n}} Y^*_{\ell m}(\hat{\mathbf{n}}) r\nabla_i \delta n^i, \\
        B^{f,r}_{\ell m} &= \frac{-i}{\sqrt{\ell(\ell+1)}} \int d\hat{\mathbf{n}} Y^*_{\ell m}(\hat{\mathbf{n}})\ \varepsilon_{ijk} n^j r \nabla_k \delta n^i.        
    \end{aligned}
\end{equation}
where, after some lengthy calculation reported in sections~\ref{subapp:explicit_calculation} and~\ref{subapp:recurvise_relation_calculation}, we obtain
\begin{equation}
    \begin{aligned}
        r\nabla_i \delta n^i &= \frac{n^j n^k e^\lambda_{jk}}{1+\hat{\mathbf{p}} \cdot \hat{\mathbf{n}}} \left[1 - \frac{1+\hat{\mathbf{p}} \cdot \hat{\mathbf{n}}}{2} e^{2\pi i f r (1+\hat{\mathbf{p}} \cdot \hat{\mathbf{n}})} \right. \\
        &\qquad\qquad\qquad\qquad \left. + \frac{i}{2\pi fr} \left( 1 + \frac{1}{1+\hat{\mathbf{p}} \cdot \hat{\mathbf{n}}} \right) \left( e^{2\pi i f r (1+\hat{\mathbf{p}} \cdot \hat{\mathbf{n}})} -1 \right) \right], \\
        \varepsilon_{ijk} n^j r \nabla_k \delta n^i &= \frac{\varepsilon_{ijk} p^i n^j n^q e^\lambda_{q k}}{1+\hat{\mathbf{p}} \cdot \hat{\mathbf{n}}} \left[ 1 - e^{2\pi i f r (1+\hat{\mathbf{p}} \cdot \hat{\mathbf{n}})} \right] . \\
    \end{aligned}
\end{equation}
The first term in square brackets corresponds to the infinite distance limit, whereas the other terms represent the finite distance correction.
In the following, we present the calculation of the harmonic coefficients in two fashions, one explicit, the other that exploits the recursive relations introduced in appendix~\ref{app:notation_conventions}.
Both procedures converge to the same results, which can be expressed in a general form as
\begin{equation}
    \begin{aligned}
        E^{f,r}_{\ell m} &= - \frac{4\pi F^E_\ell(\tilde{x})}{\sqrt{\ell(\ell+1)}} \sqrt{\frac{2\ell+1}{4\pi} \frac{(\ell-2)!}{(\ell+2)!}} \\
        &\qquad\qquad\qquad \times \int d\hat{\mathbf{p}} \left[ h_+(f,\hat{\mathbf{p}}) (\delta^K_{m,2} + \delta^K_{m,-2}) - ih_\times(f,\hat{\mathbf{p}}) (\delta^K_{m,2} - \delta^K_{m,-2}) \right], \\
        B^{f,r}_{\ell m} &= - \frac{4\pi i F^B_\ell(\tilde{x})}{\sqrt{\ell(\ell+1)}} \sqrt{\frac{2\ell+1}{4\pi} \frac{(\ell-2)!}{(\ell+2)!}} \\
        &\qquad\qquad\qquad \int d\hat{\mathbf{p}} \left[ h_\times(f,\hat{\mathbf{p}}) (\delta^K_{m,2} + \delta^K_{m,-2}) + ih_+(f,\hat{\mathbf{p}}) (\delta^K_{m,2} - \delta^K_{m,-2}) \right],
    \end{aligned}
\end{equation}
where~$\tilde{x}=2\pi fr$, and the limit~$\tilde{x}\to\infty$ for the two distance functions is~$F^{E}_\ell,F^{B}_\ell\to (-1)^\ell$.
The angular power spectrum estimators are built as
\begin{equation}
    \begin{aligned}
        \hat{C}^{EE}_\ell &= \frac{\sum_m E^{f,r}_{\ell m} \left(E^{f',r'}_{\ell m}\right)^*}{2\ell+1} \\
        &= \frac{8\pi F^E_\ell(\tilde{x}) \left[F^E_\ell(\tilde{x}')\right]^*}{\ell(\ell+1)} \frac{(\ell-2)!}{(\ell+2)!} \int d\hat{\mathbf{p}} d\hat{\mathbf{p}}' \left[ h_+(f,\hat{\mathbf{p}}) h^*_+(f',\hat{\mathbf{p}}') + h_\times(f,\hat{\mathbf{p}}) h^*_\times(f',\hat{\mathbf{p}}') \right], \\
        \hat{C}^{BB}_\ell &= \frac{\sum_m B^{f,r}_{\ell m} \left(B^{f',r'}_{\ell m}\right)^*}{2\ell+1} \\
        &= \frac{8\pi F^B_\ell(\tilde{x}) \left[F^B_\ell(\tilde{x}')\right]^*}{\ell(\ell+1)} \frac{(\ell-2)!}{(\ell+2)!} \int d\hat{\mathbf{p}} d\hat{\mathbf{p}}' \left[ h_+(f,\hat{\mathbf{p}}) h^*_+(f',\hat{\mathbf{p}}') + h_\times(f,\hat{\mathbf{p}}) h^*_\times(f',\hat{\mathbf{p}}') \right], \\
    \end{aligned}
\end{equation}
so that their expectation value reads as
\begin{equation}
    \begin{aligned}
        \left\langle \hat{C}^{EE}_\ell \right\rangle &= \frac{1}{2} \delta^D(f-f') C^{EE}_\ell(f,r,r') \\
        &= \frac{1}{2} \delta^D(f-f') \frac{8\pi}{\ell(\ell+1)} \frac{(\ell-2)!}{(\ell+2)!} F^E_\ell(2\pi fr) \left[F^E_\ell(2\pi f r')\right]^* S_h(f), \\
        \left\langle \hat{C}^{BB}_\ell \right\rangle &= \frac{1}{2} \delta^D(f-f') C^{BB}_\ell(f,r,r') \\
        &= \frac{1}{2} \delta^D(f-f') \frac{8\pi}{\ell(\ell+1)} \frac{(\ell-2)!}{(\ell+2)!} F^B_\ell(2\pi fr) \left[F^B_\ell(2\pi f r')\right]^* S_h(f).
    \end{aligned}
\end{equation}

At this stage, the expectation value can, in principle, take complex values when~$r\neq r'$.
However, if we consider angular power spectrum integrated over the frequency range, we find that
\begin{equation}
    \begin{aligned}
        C_\ell^{EE} &=
        \int df df' \left\langle \hat{C}^{EE}_\ell \right\rangle\\
        &= \frac{8\pi}{\ell(\ell+1)} \frac{(\ell-2)!}{(\ell+2)!} \int_{-\infty}^{+\infty} df \frac{1}{2} F^E_\ell (2\pi fr) \left[F^E_\ell(2\pi f r')\right]^* S_h(f) \\
        &= \frac{8\pi}{\ell(\ell+1)} \frac{(\ell-2)!}{(\ell+2)!} \int_{0}^{+\infty} df \frac{F^E_\ell (-2\pi fr) \left[F^E_\ell(-2\pi f r')\right]^* + F^E_\ell (2\pi fr) \left[F^E_\ell(2\pi f r')\right]^*}{2} S_h(f) \\
        &= \frac{8\pi}{\ell(\ell+1)} \frac{(\ell-2)!}{(\ell+2)!} \int_{0}^{+\infty} df\ \mathrm{Re}\left[F^E_\ell (2\pi fr) \left[F^E_\ell(2\pi f r')\right]^*\right] S_h(f), \\
        C_\ell^{BB} &=
        \int df df' \left\langle \hat{C}^{BB}_\ell \right\rangle \\
        &= \frac{8\pi}{\ell(\ell+1)} \frac{(\ell-2)!}{(\ell+2)!} \int_{-\infty}^{+\infty} df \frac{1}{2} F^B_\ell (2\pi fr) \left[F^B_\ell(2\pi f r')\right]^* S_h(f) \\
        &= \frac{8\pi}{\ell(\ell+1)} \frac{(\ell-2)!}{(\ell+2)!} \int_{0}^{+\infty} df \frac{F^B_\ell (-2\pi fr) \left[F^B_\ell(-2\pi f r')\right]^* + F^B_\ell (2\pi fr) \left[F^B_\ell(2\pi f r')\right]^*}{2} S_h(f) \\
        &= \frac{8\pi}{\ell(\ell+1)} \frac{(\ell-2)!}{(\ell+2)!} \int_{0}^{+\infty} df\ \mathrm{Re}\left[F^B_\ell (2\pi fr) \left[F^B_\ell(2\pi f r')\right]^*\right] S_h(f), \\
    \end{aligned}
\end{equation}
since, as we demonstrate in the next section, we have that~$\left[F^E_\ell(\tilde{x})\right]^* = F^E_\ell(-\tilde{x})$ and~$\left[F^B_\ell(\tilde{x})\right]^* = F^B_\ell(-\tilde{x})$.
Therefore, as expected, the frequency-integrated angular power spectrum is always a real quantity.

We also report for completeness the real part of the distance functions
\begin{align}
   &{\rm Re}{\left[{F^E_\ell}^*[2\pi f r]F^{E}_\ell[2\pi f r']\right]} =  \frac{1}{16 {\tilde{x}}^2 (\tilde{x}')^2} \bigg(2 {\tilde{x}} \left(\ell^2+\ell+2 {\tilde{x}}^2\right) j_{\ell-1}({\tilde{x}})\nonumber \\
   &\left[j_{\ell}({\tilde{x}'}) \left(\ell \left(\ell^3-\ell-4 (\tilde{x}')^2\right) \cos ({\tilde{x}}-{\tilde{x}'})-4 (\tilde{x}')^3 \sin ({\tilde{x}}-{\tilde{x}'})\right)
   +2 {\tilde{x}'} \left(\ell^2+\ell+2 (\tilde{x}')^2\right) j_{\ell-1}({\tilde{x}'}) \cos ({\tilde{x}}-{\tilde{x}'})\right]\nonumber\\
   &+j_{\ell}({\tilde{x}}) \left[2 {\tilde{x}'} \left(\ell^2+\ell+2 (\tilde{x}')^2\right) j_{\ell-1}({\tilde{x}'}) \left(\ell \left(\ell^3-\ell-4 {\tilde{x}}^2\right) \cos ({\tilde{x}}-{\tilde{x}'})+4 {\tilde{x}}^3 \sin ({\tilde{x}}-{\tilde{x}'})\right)\right.\nonumber\\
   &+j_{\ell}({\tilde{x}'}) \left(4 \ell \left((\tilde{x}')^3 \left(-\ell^3+\ell+4 {\tilde{x}}^2\right)+\ell \left(\ell^2-1\right) {\tilde{x}}^3-4 {\tilde{x}}^3 (\tilde{x}')^2\right) \sin ({\tilde{x}}-{\tilde{x}'})\right.\nonumber\\
   &\left.\left.+\left(4 \ell^2 (\tilde{x}')^2 \left(-\ell^3+\ell+4 {\tilde{x}}^2\right)
   +\left(\ell^2-1\right) \ell^3 \left(\ell^3-\ell-4 {\tilde{x}}^2\right)+16 {\tilde{x}}^3 (\tilde{x}')^3\right) \cos ({\tilde{x}}-{\tilde{x}'})\right)\right]\bigg)
\end{align}
and
\begin{align}
    &{\rm Re}{\left[{F^B_\ell}^*[2\pi f r]F^{B}_\ell[2\pi f r']\right]} = \nonumber\\
    &\frac{1}{4 {\tilde{x}} {\tilde{x}'}}j_\ell({\tilde{x}}) \left[j_\ell({\tilde{x}'}) \left(\left(2 (\tilde{x}')^2 \left(\ell^3-\ell+2 {\tilde{x}}^2\right)+(\ell-1)^2 \ell^2 {\tilde{x}} {\tilde{x}'}+\left(\ell^2-1\right) \ell \left(\ell^3-\ell+2 {\tilde{x}}^2\right)\right) \cos ({\tilde{x}}-{\tilde{x}'})\right.\right.\nonumber\\
    &\left.-(\ell-1) \ell ({\tilde{x}}-{\tilde{x}'}) \left(\ell^3-\ell-2 {\tilde{x}} {\tilde{x}'}\right) \sin ({\tilde{x}}-{\tilde{x}'})\right)\nonumber\\
    &+{\tilde{x}'} j_{\ell-1}({\tilde{x}'}) \left(\left(2 {\tilde{x}'} \left(\ell^3-\ell+2 {\tilde{x}}^2\right)+\left(\ell^2-1\right) \ell^2 {\tilde{x}}\right) \sin ({\tilde{x}}-{\tilde{x}'})\right.\nonumber\\
    &\left.\left.-\ell \left((\ell+1) \left(\ell^3-\ell+2 {\tilde{x}}^2\right)-2 (\ell-1) {\tilde{x}} {\tilde{x}'}\right) \cos ({\tilde{x}}-{\tilde{x}'})\right)\right]\nonumber\\
    &+\frac{1}{4 {\tilde{x}'}} \left[j_{\ell-1}({\tilde{x}}) \left({\tilde{x}'} j_{\ell-1}({\tilde{x}'}) \left(\left(\ell^2 (\ell+1)^2+4 {\tilde{x}} {\tilde{x}'}\right) \cos ({\tilde{x}}-{\tilde{x}'})+2 \ell (\ell+1) ({\tilde{x}}-{\tilde{x}'}) \sin ({\tilde{x}}-{\tilde{x}'})\right)\right.\right.\nonumber\\
    &-j_\ell({\tilde{x}'}) \left(\left(2 \left(\ell^2-1\right) \ell {\tilde{x}}+\left(\ell^2-1\right) \ell^2 {\tilde{x}'}+4 {\tilde{x}} x(\tilde{x}')^2\right) \sin ({\tilde{x}}-{\tilde{x}'})\right.\nonumber\\
    &\left.\left.\left.+\ell \left(-2 (\ell-1) {\tilde{x}} {\tilde{x}'}+2 (\ell+1) (\tilde{x}')^2+(\ell-1) \ell (\ell+1)^2\right) \cos ({\tilde{x}}-{\tilde{x}'})\right)\right)\right]\,.
\end{align}

\begin{figure}[ht]
    \centerline{
    \includegraphics[width=\linewidth]{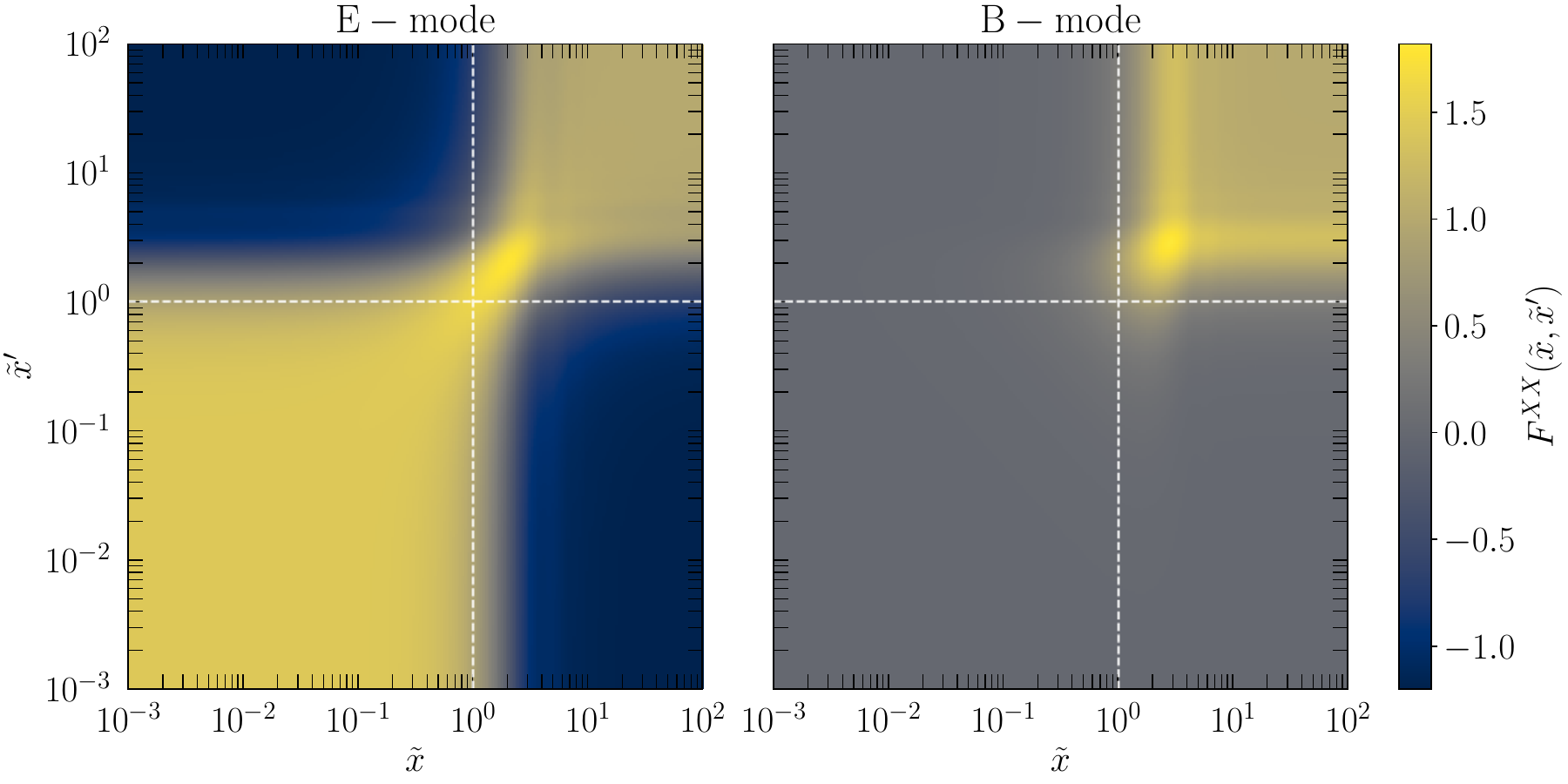}}
    \caption{Value of the real part of the E-mode (\textit{left panel}) and B-mode (\textit{right panel}) form factors for~$\ell=2$ and pairs of sources located at generic~$\tilde{x}$ and~$\tilde{x}'$.
    The \textit{dashed white lines} correspond to~$\tilde{x} = \tilde{x}' = 1$.}
    \label{fig::cross_x_xp}
\end{figure}

We show in figure~\ref{fig::cross_x_xp} the values of the E- and B-mode form factors for~$\ell=2$ and different combinations of~$\tilde{x}=2\pi f r$ and~$\tilde{x}'=2\pi f r'$.
In general, we observe that the E-mode form factor is positive only when both sources are either in the short- or in the long-distance limit.
On the other hand, the B-mode form factor is non-vanishing only when both sources are in the long-distance limit, as we also observed from figure~\ref{fig:FEB2_of_x}.
Moreover, in both panels we can observe the oscillatory features of the form factors at~$\tilde{x}\approx \tilde{x}' \approx \mathcal{O}(\mathrm{few})$, although it is not very prominent for~$\ell=2$.


\subsubsection{Explicit calculation}
\label{subapp:explicit_calculation}

Here we want to explicitly compute the E- and B-mode harmonic coefficient, which have the generic structure given by
\begin{equation}
    X_{\ell m} = \int d\hat{\mathbf{n}} X(\hat{\mathbf{n}}) Y^*_{\ell m} (\hat{\mathbf{n}}) = \sqrt{\frac{2\ell+1}{4\pi}\frac{(\ell-m)!}{(\ell+m)!}} \int d\mu d\phi X(\mu,\phi) P_\ell^m(\mu) e^{-im\phi}.
\end{equation} 

Starting from the integrand of equation~\eqref{eq::dn_full}, we define
\begin{equation}
    \begin{aligned}
        & \mathcal{A}^i = \frac{n^i n^j n^k e^\lambda_{jk}}{2(1+\gamma)}\,,\quad \mathcal{B}^i = \frac{p^i n^j n^k e^\lambda_{jk}}{2(1+\gamma)}\,,\quad \mathcal{C}^i = n_j e^\lambda_{ji}\,, \\
        & a=1-\frac{i}{\tilde{x}}\frac{(2+\gamma)}{(1+\gamma)}[1-e^{i\tilde{x}(1+\gamma)}]\,,\quad b = 1 - \frac{i}{\tilde{x}(1+\gamma)} [1-e^{i\tilde{x}(1+\gamma)}]\,,\quad c = b-\frac{1}{2}\,,
    \end{aligned}
\end{equation}
where~$\gamma = \hat{\mathbf{p}}\cdot\hat{\mathbf{n}}$. 
The angular gradient of these quantities is given by
\begin{equation}
    \begin{aligned}
        \nabla_{\Omega,j} \mathcal{A}^i = & \frac{[n_mn_k\delta_{ij}+\delta_{mj}n_in_k+\delta_{kj}n_in_m-3n_kn_jn_in_m](1+\gamma)}{2(1+\gamma)^2} e^\lambda_{mk} \\
        &\quad -\frac{n^in^mn^kp_j+\gamma n^in^mn^kn_j}{2(1+\gamma)^2} e^\lambda_{mk}, \\
        \nabla_{\Omega,j} \mathcal{B}^i = & \left[\frac{p_i \delta_{mj}n_k}{(1+\gamma)}-\frac{n_mn_k}{2(1+\gamma)^2}(2n_jp_i+n_jp_i\gamma+p_ip_j)\right], \\
        \nabla_{\Omega,j} \mathcal{C}^i = &\, e^\lambda_{ij}-n_jn_k e^\lambda_{ik}, \\
        \nabla_{\Omega,j} a = &\left[\frac{i}{\tilde{x}} \frac{[1-e^{i\tilde{x}(1+\gamma)}]}{(1+\gamma)^2}-\frac{(2+\gamma)}{(1+\gamma)}e^{i\tilde{x}(1+\gamma)}\right] \nabla_{\Omega,j} \gamma, \\
        \nabla_{\Omega,j} b = \nabla_{\Omega,j} c = &\left[\frac{i}{\tilde{x}} \frac{[1-e^{i\tilde{x}(1+\gamma)}]}{(1+\gamma)^2}-\frac{e^{i\tilde{x}(1+\gamma)}}{(1+\gamma)}\right] \nabla_{\Omega,j}\gamma.
    \end{aligned}
\end{equation}

In the case of the E-mode harmonic coefficient, we have to consider the divergence of the angular deflection, thus
\begin{equation}
    \begin{aligned}
        \nabla_{\Omega,j} \delta n^j = &\nabla_{\Omega,j}  [a\mathcal{A}^j+b\mathcal{B}^j-c\mathcal{C}^j] \\
        = & \left[1-\frac{i}{\tilde{x}}\frac{(2+\gamma)}{(1+\gamma)}+\left[\frac{i}{\tilde{x}}\frac{(2+\gamma)}{(1+\gamma)}-\frac{1}{2}(1+\gamma)\right]e^{i\tilde{x}(1+\gamma)}\right]\frac{n_m n_k e^\lambda_{mk}}{1+\gamma}.
    \end{aligned}
\end{equation}

Since it is always possible to choose a reference frame where~$\hat{\mathbf{p}}$ is oriented along the z-axis, we have that~$\gamma=\mu$.
Therefore, we obtain the results showed in equation~\eqref{eq::tens_proj}, and the integrals over the angle~$\phi$ are those reported in equation~\eqref{eq::int_phi}.
Thus, the only remaining integral to be solved is
\begin{align}
\label{Eq::Full_E_integral}
    \int_{-1}^1 d\mu P_\ell^{m = 2}(\mu) (1-\mu) \left[1-\frac{i}{\tilde{x}}\frac{(2+\mu)}{(1+\mu)}+\left[\frac{i}{\tilde{x}}\frac{(2+\mu)}{(1+\mu)}-\frac{1}{2}(1+\mu)\right]e^{i\tilde{x}(1+\mu)}\right]\,.
\end{align}
Using the properties of Legendre polynomials, we obtain
\begin{align}
    \int_{-1}^1 d\mu P_\ell^{m = 2}(\mu) (1-\mu)= & 4 (-1)^\ell, \\
    \int_{-1}^1 d\mu P_\ell^{m = 2}(\mu)(1-\mu)\frac{(2+\mu)}{(1+\mu)} = &\int_{-1}^1 d \mu (1-\mu)^2(2+\mu) \frac{d^2 P_\ell(\mu)}{d\mu^2}\nonumber\\
    = &\left[(1-\mu)^2 (2+\mu)\frac{d P_\ell(\mu)}{d\mu}\right]_{-1}^1 - \int_{-1}^1 3 [\mu^2-1]\frac{dP_\ell (\mu)}{d\mu} \nonumber\\
    = & 2 \ell(\ell+1)(-1)^\ell, \\
    \int_{-1}^1 d\mu P_{\ell}^{m =2}(\mu)(1-\mu^2) e^{i\tilde{x}(1+\mu)} = & 2j_{\ell}(\tilde{x})i^\ell e^{i\tilde{x}}\left[\frac{(1+\ell)(2+\ell)(2\ell^2+6\ell-3)}{(2\ell+3)(2\ell+1)-(\ell+1)(\ell+2)}\right] \nonumber\\
    &-2 \frac{\ell(\ell-1)(\ell+1)(\ell+2)}{(2\ell+1)(2\ell+3)}j_{\ell+2}(\tilde{x})i^\ell e^{i\tilde{x}}\nonumber\\
    &- 2\frac{\ell(\ell-1)(\ell+1)(\ell+2)}{(2\ell-1)(2\ell+1)}j_{\ell-2}(\tilde{x})i^\ell e^{i\tilde{x}}, \\
    \int_{-1}^1P_{\ell}^{m=2}(1-\mu)\frac{(2+\mu)}{(1+\mu)} e^{i\tilde{x}(1+\mu)} = & e^{i\tilde{x}} \int_{-1}^1 d\mu P_{\ell}^{m=2}\left(\frac{2}{1+\mu}-\mu\right)e^{i\tilde{x}\mu}\nonumber\\
    = &4(-1)^\ell i\tilde{x} + 2(-1)^\ell\ell(\ell+1)-4e^{i\tilde{x}}j_{\ell}(\tilde{x})\tilde{x}i^\ell[\tilde{x}+i(\ell+1)]\nonumber\\
    &+4\tilde{x}^2i i^\ell j_{\ell+1}(\tilde{x})e^{i\tilde{x}}+\frac{2\ell(\ell-1)(\ell+1)}{(2\ell+1)}e^{i\tilde{x}}i^{\ell+1}j_{\ell+1}(\tilde{x})\nonumber\\
    & +\frac{2\ell(1+\ell)(2+\ell)}{2\ell+1}e^{i\tilde{x}}i^{\ell-1}j_{\ell-1}(\tilde{x}).
\end{align}
Moreover, thanks to the properties of the associated Legendre polynomials described in appendix~\ref{app:notation_conventions}, the calculation for~$P_{\ell}^{m=-2}$ returns the same results of that of~$P_{\ell}^{m=2}$.
In the end, using the recursive relations of the Bessel functions, $j_{\ell}(\tilde{x})=\frac{(2\ell-1)}{\tilde{x}}j_{\ell-1}(\tilde{x})-j_{\ell-2}(\tilde{x})$, we find
\begin{equation}
    \begin{aligned}
        E^{f,r}_{\ell m} = & -\frac{4\pi}{\sqrt{\ell(\ell+1)}} \sqrt{\frac{2\ell+1}{4\pi}\frac{(\ell-2)!}{(\ell+2)!}}\int d\hat{\mathbf{p}}\left[h^+_{f,\hat{\mathbf{p}}}(\delta_{m,2}+\delta_{m,-2})-ih^{\times}_{f,\hat{\mathbf{p}}}(\delta_{m,2}-\delta_{m,-2})\right]\\
        & \times\left\{e^{i\tilde{x}} i^\ell \left[ j_{\ell-1}(\tilde{x})\left(\frac{\ell(\ell+1)}{2\tilde{x}}+\tilde{x}\right)-j_\ell(\tilde{x})\left(i\tilde{x}-\frac{\ell^2(\ell^2-1)}{4\tilde{x}^2}+\ell\right)\right]\right\}\,.
    \end{aligned}
\end{equation}

On the other hand, in the case of B-modes, we have
\begin{equation}
    (\hat{\mathbf{n}} \times \nabla^{(2)})_i \delta n^i = \varepsilon_{kij} n^i \nabla_{\Omega,j} \delta n^k,
\end{equation}
which implies
\begin{equation}
    B(\hat{\mathbf{n}}) = \varepsilon_{ikj} n^k\nabla_{\Omega,j}\delta n^i = \varepsilon_{ikj} n^k \nabla_{\Omega, j} [a\mathcal{A}^i+b\mathcal{B}^i-c\mathcal{C}^i]\,.
\end{equation}
Using the results obtained above, and observing that the non-symmetric parts of~$n^k\nabla_{\Omega, j}\mathcal{A}^i$ and~$n^k\nabla_{\Omega,j} \mathcal{C}^i$ vanish, we get
\begin{align}
    B(\hat{\mathbf{n}})
    = & \varepsilon_{ikj} \left[\frac{n^k}{(1+\gamma)} \left(p_i e^\lambda_{j\ell}n_{\ell}+n_m e^\lambda_{mi} p_j e^{i\tilde{x}(1+\gamma)}\right)-\frac{i}{\tilde{x}} \frac{[1-e^{i\tilde{x}(1+\gamma)}]}{(1+\gamma)^2} n^k (p_i e^\lambda_{j\ell}n_{\ell}+n_m e^\lambda_{mi}p_j) \right],
\end{align}
which can be further simplified by observing that
\begin{align}
    \varepsilon_{ikj}\left(p_i e^\lambda_{j\ell}n_{\ell}+n_m e^\lambda_{mi}p_je^{i\tilde{x}(1+\gamma)}\right) = & \varepsilon_{ikj}p_i e^\lambda_{j\ell}n_{\ell}+\varepsilon_{jki}n_m e^\lambda_{mj}p_ie^{i\tilde{x}(1+\gamma)}\nonumber\\
    = & \varepsilon_{ikj}p_i e^\lambda_{j\ell}n_{\ell}\left[1-e^{i\tilde{x}(1+\gamma)}\right], \\
    \varepsilon_{ikj}(p_i e^\lambda_{j\ell}n_{\ell}+n_m e^\lambda_{mi}p_j)= & \varepsilon_{ikj}(p_i e^\lambda_{j\ell}n_{\ell}-n_m e^\lambda_{mj}p_i) = 0\,,
\end{align}
yielding
\begin{align}
    B(\hat{\mathbf{n}}) = \varepsilon_{ikj} \frac{p_i e^\lambda_{j\ell}n_{\ell}n_k}{(1+\gamma)}\left[1-e^{i\tilde{x}(1+\gamma)}\right].
\end{align}
Therefore, in this case, we need to solve the two additional integrals
\begin{align}
    \int_{-1}^1d\mu P_{\ell}^{m=2}(\mu)e^{i\tilde{x}\mu} = & 2[e^{i\tilde{x}}+(-1)^\ell e^{-i\tilde{x}}] + 4\tilde{x}i^\ell j_{\ell+1}(\tilde{x})-2(\ell+1)(\ell+2)i^\ell j_\ell(\tilde{x})\\
    \int_{-1}^1d\mu P_{\ell}^{m=2}(\mu)\mu e^{i\tilde{x}\mu} = & 2[e^{i\tilde{x}}-(-1)^\ell e^{-i\tilde{x}}]-(i\tilde{x}) 4 i^\ell j_{\ell}(\tilde{x}) - \frac{2\ell(\ell-1)(\ell+1)}{(2\ell+1)} i^{\ell+1}j_{\ell+1}(\tilde{x})\nonumber\\
& - \frac{2\ell(1+\ell)(2+\ell)}{1+2\ell}i^{\ell-1}j_{\ell-1}(\tilde{x})\,.
\end{align}
In the end, the B-mode harmonic coefficients read as
\begin{equation}
    \begin{aligned}
        B^{f,r}_{\ell m}  = &-\frac{4\pi i}{\sqrt{\ell(\ell+1)}} \sqrt{\frac{2\ell+1}{4\pi}\frac{(\ell-2)!}{(\ell+2)!}}\int d\hat{\mathbf{p}}\left[h^{\times}_{f,\hat{\mathbf{p}}}(\delta_{m,2}+\delta_{m,-2})+ih^{+}_{f,\hat{\mathbf{p}}}(\delta_{m,2}-\delta_{m,-2})\right]\\
        & \times (-i) i^\ell e^{i\tilde{x}}\left[j_\ell(\tilde{x})\left(\frac{\ell(\ell-1)}{2}i+\frac{\ell(\ell^2-1)}{2\tilde{x}}+\tilde{x}\right)+j_{\ell-1}(\tilde{x})\left(i\tilde{x}-\frac{\ell(1+\ell)}{2}\right)\right]\,.
    \end{aligned}
\end{equation}

\underline{\textbf{Parity of the form functions.}}
Finally, in this small section, we show how these new form factors change under a frequency reversal transformation~$f\to -f$, or, more generally, $x\to -x$.
Under such a coordinate change, we have that
\begin{equation}
    \begin{aligned}
        F^E_\ell (-\tilde{x}) &= i^\ell e^{i(-\tilde{x})} \left[ j_{\ell-1}(-\tilde{x}) \left(\frac{\ell(\ell+1)}{2(-\tilde{x})} + (-\tilde{x}) \right) - j_\ell(-\tilde{x}) \left( i(-\tilde{x}) - \frac{\ell^2(\ell^2-1)}{4(-\tilde{x})^2} + \ell\right)\right] \\
        &= (-i)^\ell e^{-i\tilde{x}} \left[ j_{\ell-1}(\tilde{x}) \left(\frac{\ell(\ell+1)}{2\tilde{x}} + \tilde{x} \right) - j_\ell(\tilde{x}) \left( -i\tilde{x} - \frac{\ell^2(\ell^2-1)}{4\tilde{x}^2} + \ell \right) \right] \\
        &= \left[F^E_\ell(\tilde{x})\right]^*, \\
        F^B_\ell (-\tilde{x}) &= (-i) i^\ell e^{i(-\tilde{x})} \left[ j_\ell(-\tilde{x}) \left(i \frac{\ell(\ell-1)}{2} + \frac{\ell(\ell^2-1)}{2(-\tilde{x})} + (-\tilde{x}) \right) + j_{\ell-1}(-\tilde{x}) \left( i(-\tilde{x}) - \frac{\ell(\ell+1)}{2} \right) \right] \\
        &= i (-i)^\ell e^{-i\tilde{x}} \left[ j_\ell(\tilde{x}) \left(-i \frac{\ell(\ell-1)}{2} + \frac{\ell(\ell^2-1)}{2\tilde{x}} + \tilde{x} \right) + j_{\ell-1}(\tilde{x}) \left( -i\tilde{x} - \frac{\ell(\ell+1)}{2} \right) \right] \\
        &= \left[F^B_\ell(\tilde{x})\right]^*,
    \end{aligned}
\end{equation}
where we used the~$j_\ell(-\tilde{x}) = (-1)^\ell j_\ell(\tilde{x})$ property of the spherical Bessel functions.


\subsubsection{Calculation via recursive relations}
\label{subapp:recurvise_relation_calculation}

The results of the previous sections can also be derived via the recursive relations reported in appendix~\ref{app:notation_conventions}.
The E-mode harmonic coefficients for the two polarization degrees of freedom read as
\begin{equation}
    \begin{aligned}
        E^{f,r,+}_{\ell m} &= -\frac{1}{\sqrt{\ell(\ell+1)}} \int d\hat{\mathbf{p}} h_+ (f,\hat{\mathbf{p}}) \int d\hat{\mathbf{n}} Y^*_{\ell m}(\hat{\mathbf{n}}) \frac{n^j n^k e^+_{j k}}{1+\hat{\mathbf{p}} \cdot \hat{\mathbf{n}}} \\
        &\qquad \times \left[1 - \frac{1+\hat{\mathbf{p}} \cdot \hat{\mathbf{n}}}{2} e^{2\pi i f r (1+\hat{\mathbf{p}} \cdot \hat{\mathbf{n}})} + \frac{i}{2\pi fr} \left( 1 + \frac{1}{1+\hat{\mathbf{p}} \cdot \hat{\mathbf{n}}} \right) \left( e^{2\pi i f r (1+\hat{\mathbf{p}} \cdot \hat{\mathbf{n}})} -1 \right) \right]  \\
        &= -\frac{\pi}{\sqrt{\ell(\ell+1)}} \sqrt{\frac{2\ell+1}{4\pi} \frac{(\ell-2)!}{(\ell+2)!} } \int d\hat{\mathbf{p}} h_+ (f,\hat{\mathbf{p}}) ( \delta^K_{m,2} + \delta^K_{m,-2}) \\
        &\quad \times \int dy (1-y)(1-y^2) \left[1 - \frac{1+y}{2}e^{ix(1+y)} + \frac{i}{x} \left(1+ \frac{1}{1+y} \right) \left( e^{ix(1+y)} - 1 \right) \right] \frac{d^2P_\ell}{dy^2} \\
        &= -\frac{\pi}{\sqrt{\ell(\ell+1)}} \sqrt{\frac{2\ell+1}{4\pi} \frac{(\ell-2)!}{(\ell+2)!} } \int d\hat{\mathbf{p}} h_+ (f,\hat{\mathbf{p}}) ( \delta^K_{m,2} + \delta^K_{m,-2}) \\
        &\qquad\qquad \times \left[(J_0-J_1-J_2+J_3) - \frac{1}{2}e^{ix}(K_0-2K_2+K_4) \right. \\
        &\qquad\qquad\qquad\qquad \left. - \frac{i}{x} (2J_0-3J_1+J_3) + \frac{i}{x}e^{ix} (2K_0 - 3K_1 + K_3) \right] \\
        &= -\frac{4\pi }{\sqrt{\ell(\ell+1)}} \sqrt{\frac{2\ell+1}{4\pi} \frac{(\ell-2)!}{(\ell+2)!} } \int d\hat{\mathbf{p}} h_+ (f,\hat{\mathbf{p}}) ( \delta^K_{m,2} + \delta^K_{m,-2}) \\
        &\qquad\qquad \times \frac{1}{4} i^\ell e^{ix} \left[ (16-4ix+x^2) j_\ell + \frac{12+14x^2}{x} \partial_x j_\ell \right. \\
        &\qquad\qquad\qquad\qquad\qquad \left. + (24+2x^2) \partial^2_x j_\ell + 10x\partial^3_x j_\ell + x^2 \partial^4_x j_\ell \right] \\
        E^{f,r,\times}_{\ell m} &= -\frac{4\pi }{\sqrt{\ell(\ell+1)}} \sqrt{\frac{2\ell+1}{4\pi} \frac{(\ell-2)!}{(\ell+2)!} } \int d\hat{\mathbf{p}}\ (-i) h_\times (f,\hat{\mathbf{p}}) ( \delta^K_{m,2} - \delta^K_{m,-2}) \\
        &\qquad\qquad \times \frac{1}{4} i^\ell e^{ix} \left[ (16-4ix+x^2) j_\ell + \frac{12+14x^2}{x} \partial_x j_\ell \right. \\
        &\qquad\qquad\qquad\qquad\qquad \left. + (24+2x^2) \partial^2_x j_\ell + 10x\partial^3_x j_\ell + x^2 \partial^4_x j_\ell \right] . \\
    \end{aligned}
\end{equation}
On the other hand, the B-mode harmonic coefficients are given by
\begin{equation}
    \begin{aligned}
        B^{f,r,+}_{\ell m} &= - \frac{i}{\sqrt{\ell(\ell+1)}} \int d\hat{\mathbf{p}} h_+(f,\hat{\mathbf{p}}) \int d\hat{\mathbf{n}} Y^*_{\ell m}(\hat{\mathbf{n}}) \frac{\varepsilon_{ijk} p^i n^j n^q e^+_{q k}}{1+\hat{\mathbf{p}} \cdot \hat{\mathbf{n}}} \left[ 1 - e^{2\pi i f r (1+\hat{\mathbf{p}} \cdot \hat{\mathbf{n}})} \right] \\
        &= - \frac{\pi i}{\sqrt{\ell(\ell+1)}} \sqrt{\frac{2\ell+1}{4\pi} \frac{(\ell-2)!}{(\ell+2)!} } \int d\hat{\mathbf{p}}\ ih_+(f,\hat{\mathbf{p}}) ( \delta^K_{m,2} - \delta^K_{m,-2}) \\
        &\qquad\qquad \times \int dy (1-y) (1-y^2) \left[ 1-e^{ix(1+y)} \right] \frac{d^2P_\ell}{dy^2} \\
        &= - \frac{\pi i}{\sqrt{\ell(\ell+1)}} \sqrt{\frac{2\ell+1}{4\pi} \frac{(\ell-2)!}{(\ell+2)!} } \int d\hat{\mathbf{p}}\ ih_+(f,\hat{\mathbf{p}}) ( \delta^K_{m,2} - \delta^K_{m,-2}) \\
        &\qquad\qquad \times \left[ \left(J_0 - J_1 - J_2 + J_3\right) - e^{ix} \left( K_0 - K_1 - K_2 + K_3 \right) \right] \\
        &= - \frac{4 \pi i }{\sqrt{\ell(\ell+1)}} \sqrt{\frac{2\ell+1}{4\pi} \frac{(\ell-2)!}{(\ell+2)!} } \int d\hat{\mathbf{p}}\ ih_+(f,\hat{\mathbf{p}}) ( \delta^K_{m,2} - \delta^K_{m,-2}) \\
        &\qquad \times \frac{1}{2} i^\ell e^{ix} \left[ (2+2ix+x^2) j_\ell + (6i+4x+ix^2) \partial_x j_\ell + x(6i+x) \partial^2_x j_\ell + ix^2 \partial^3_x j_\ell \right], \\
        B^{f,r,\times}_{\ell m} &= - \frac{4 \pi i}{\sqrt{\ell(\ell+1)}} \sqrt{\frac{2\ell+1}{4\pi} \frac{(\ell-2)!}{(\ell+2)!} } \int d\hat{\mathbf{p}}\ h_\times (f,\hat{\mathbf{p}}) ( \delta^K_{m,2} + \delta^K_{m,-2}) \\
        &\qquad \times \frac{1}{2} i^\ell e^{ix} \left[ (2+2ix+x^2) j_\ell + (6i+4x+ix^2) \partial_x j_\ell + x(6i+x) \partial^2_x j_\ell + ix^2 \partial^3_x j_\ell \right]. \\
    \end{aligned}
\end{equation}
The form factors found in this approach exactly match those of the previous section.
The interested reader can easily prove it by using the spherical Bessel function recursive relations and the fact that
\begin{equation}
    \begin{aligned}
        \partial_x j_\ell &= \frac{\ell}{x} j_\ell - j_{\ell+1}, \\
        \partial^2_x j_\ell &= \frac{\ell(\ell-1) - x^2}{x^2} j_\ell + \frac{2}{x} j_{\ell+1}, \\
        \partial^3_x j_\ell &= \frac{(\ell-2) \left[\ell(\ell-1) - x^2 \right]}{x^3} j_\ell + \frac{x^2 - \ell(\ell+1) - 6}{x^2} j_{\ell+1}, \\
        \partial^4_x j_\ell &= \frac{x^4 + \ell(\ell-1)(\ell-2)(\ell-3) - 2x^2 \left[ \ell(\ell-1) + 4 \right]}{x^4} j_\ell + \frac{4 \left[ 2\ell(\ell+1) - x^2 + 6 \right]}{x^3} j_{\ell+1}. \\
    \end{aligned}
\end{equation}


\subsection{Redshift angular power spectrum}

The redshift harmonic coefficients for the two GR polarization degrees of freedom are given by
\begin{equation}
    \begin{aligned}
        z^{f,+}_\mathrm{\ell m} &= \int d\hat{\mathbf{n}}\ z^+(f,\hat{\mathbf{n}}) Y^*_{\ell m}(\hat{\mathbf{n}}) = \int d\hat{\mathbf{p}} h_+(f,\hat{\mathbf{p}}) \int d\hat{\mathbf{n}} \frac{n^j n^k e^+_{jk}}{2(1+\hat{\mathbf{p}}\cdot\hat{\mathbf{n}})} Y^*_{\ell m}(\hat{\mathbf{n}}) \\
        &= \frac{1}{2} \sqrt{\frac{2\ell+1}{4\pi} \frac{(\ell-m)!}{(\ell+m)!}} \int d\hat{\mathbf{p}}\ h_+(f,\hat{\mathbf{p}}) \int d\cos\theta (1-\cos\theta) P_\ell^m(\cos\theta) \int d\phi \cos(2\phi) e^{-im\phi} \\
        &= 2\pi (-1)^\ell \sqrt{\frac{2\ell+1}{4\pi} \frac{(\ell-2)!}{(\ell+2)!}} \int d\hat{\mathbf{p}}\ h_+(f,\hat{\mathbf{p}}) (\delta^K_{m,2} + \delta^K_{m,-2}) \\
        z^{f,\times}_\mathrm{\ell m} &= 2\pi (-1)^\ell \sqrt{\frac{2\ell+1}{4\pi} \frac{(\ell-2)!}{(\ell+2)!}} \int d\hat{\mathbf{p}}\ (-i)h_\times(f,\hat{\mathbf{p}}) (\delta^K_{m,2} - \delta^K_{m,-2}) \\
    \end{aligned}
\end{equation}
using the results of the appendix~\ref{app:notation_conventions}.
The estimator of the angular power spectrum of~$z^f_\mathrm{\ell m} = z^{f,+}_\mathrm{\ell m}+z^{f,\times}_\mathrm{\ell m}$ is then built as
\begin{equation}
    \hat{C}^{zz}_\ell = \frac{\sum_m z^f_\mathrm{\ell m} \left(z^{f'}_\mathrm{\ell m} \right)^*}{2\ell+1} = 2\pi \frac{(\ell-2)!}{(\ell+2)!} \int d\hat{\mathbf{p}} d\hat{\mathbf{p}} \left[ h_+(f,\hat{\mathbf{p}}) h^*_+(f',\hat{\mathbf{p}}') + h_\times(f,\hat{\mathbf{p}}) h^*_\times(f',\hat{\mathbf{p}}') \right],
\end{equation}
and its expectation value is given by
\begin{equation}
    \left\langle \hat{C}^{zz}_\ell \right\rangle = \frac{1}{2} \delta^D(f-f') C^{zz}_\ell(f) = \frac{1}{2} \delta^D(f-f') 4\pi \frac{(\ell-2)!}{(\ell+2)!} S_h(f).
\end{equation}

Alternatively, the two-point function in harmonic space can also be directly computed from the angle-dependent two-point function as
\begin{equation}
    \begin{aligned}
        \left\langle z^f_{\ell m} z^{f'}_{\ell' m'} \right\rangle &= \frac{1}{2}\delta^D(f-f') S_h(f) \int d\hat{\mathbf{n}} d\hat{\mathbf{n}}' Y^*_{\ell m}(\hat{\mathbf{n}}) Y_{\ell' m'}(\hat{\mathbf{n}}') \mathrm{HD}(\hat{\mathbf{n}}\cdot\hat{\mathbf{n}}') \\
        &= \frac{1}{2}\delta^D(f-f') S_h(f) \int d\hat{\mathbf{n}} d\hat{\mathbf{n}}' Y^*_{\ell m}(\hat{\mathbf{n}}) Y_{\ell' m'}(\hat{\mathbf{n}}') \sum_{\ell''} \gamma_{\ell''} P_{\ell''}(\hat{\mathbf{n}}\cdot\hat{\mathbf{n}}') \\
        &= \frac{1}{2}\delta^D(f-f') S_h(f) \int d\hat{\mathbf{n}} d\hat{\mathbf{n}}' Y^*_{\ell m}(\hat{\mathbf{n}}) Y_{\ell' m'}(\hat{\mathbf{n}}') \sum_{\ell'' m''} \gamma_{\ell''} \frac{4\pi}{2\ell''+1} Y_{\ell'' m''}(\hat{\mathbf{n}}) Y^*_{\ell'' m''}(\hat{\mathbf{n}}') \\
        &= \frac{1}{2}\delta^D(f-f') \delta^K_{\ell\ell'} \delta^K_{mm'} \frac{4\pi}{2\ell+1} \gamma_\ell S_h(f) = \frac{1}{2}\delta^D(f-f') \delta^K_{\ell\ell'} \delta^K_{mm'} C^{zz}_\ell(f),
    \end{aligned}
\end{equation}
where the Legendre coefficients~$\gamma_\ell$ of the Hellings-Downs curve, given by~\cite{Qin:2018yhy}
\begin{equation}
    \gamma_0 = \gamma_1 = 0, \quad \gamma_\ell = (2\ell+1) \frac{(\ell-2)!}{(\ell+2)!},
\end{equation}
ensure that we recover the result presented above.


\subsection{Cross-correlation of observables}

So far we provided a form only for the auto-angular power spectrum; however redshift and angular displacement are in principle correlated, since both effects are generated by the same GWB.
Therefore, it is possible to built the estimator for the cross-angular power spectra as
\begin{equation}
    \begin{aligned}
        \hat{C}^{Ez}_\ell &= \frac{\sum_m E^{f,r}_{\ell m} \left( z^{f'}_{\ell m} \right)^*}{2\ell+1} \\
        &= -\frac{2\pi F^E_\ell (x)}{\sqrt{\ell(\ell+1)}} \frac{(\ell-2)!}{(\ell+2)!} \int d\hat{\mathbf{p}} d\hat{\mathbf{p}}' \left[ h_+(f,\hat{\mathbf{p}}) h^*_+(f',\hat{\mathbf{p}}') + h_\times(f,\hat{\mathbf{p}}) h^*_\times(f',\hat{\mathbf{p}}') \right], \\
        \hat{C}^{Bz}_\ell &= \frac{\sum_m B^{f,r}_{\ell m} \left( z^{f'}_{\ell m} \right)^*}{2\ell+1} \\
        &= -\frac{2\pi i F^B_\ell (x)}{\sqrt{\ell(\ell+1)}} \frac{(\ell-2)!}{(\ell+2)!} \int d\hat{\mathbf{p}} d\hat{\mathbf{p}}' \left[ h_+(f,\hat{\mathbf{p}}) h^*_\times(f',\hat{\mathbf{p}}') + h_\times(f,\hat{\mathbf{p}}) h^*_+(f',\hat{\mathbf{p}}') \right], \\
        \hat{C}^{EB}_\ell &= \frac{\sum_m E^{f,r}_{\ell m} \left( B^{f',r'}_{\ell m} \right)^*}{2\ell+1} \\
        &= -\frac{4\pi i F^E_\ell (x) \left[ F^B_\ell(x') \right]^*}{\ell(\ell+1)} \frac{(\ell-2)!}{(\ell+2)!} \int d\hat{\mathbf{p}} d\hat{\mathbf{p}}' \left[ h_+(f,\hat{\mathbf{p}}) h^*_\times(f',\hat{\mathbf{p}}') + h_\times(f,\hat{\mathbf{p}}) h^*_+(f',\hat{\mathbf{p}}') \right], \\
    \end{aligned}
\end{equation}
therefore, their expectation value is given by
\begin{equation}
    \begin{aligned}
        \left\langle \hat{C}^{Ez}_\ell \right\rangle &= \frac{1}{2} \delta^D(f-f') C^{Ez}_\ell(f,r) = \frac{1}{2} \delta^D(f-f') \times \left[ -\frac{4\pi F^E_\ell (\tilde{x})}{\sqrt{\ell(\ell+1)}} \frac{(\ell-2)!}{(\ell+2)!} S_h(f) \right], \\
        \left\langle \hat{C}^{Bz}_\ell \right\rangle &= \left\langle \hat{C}^{EB}_\ell \right\rangle = 0,
    \end{aligned}
\end{equation}
where the absence of correlation between the components~$(E,z)$ and~$B$ is due to the statistical independence of the polarization degrees of freedom.
As previously commented, although at this stage~$C^{Ez}_\ell$ is still a complex quantity, once we integrate over frequencies we obtain that
\begin{equation}
    \begin{aligned}
        \int df df' \left\langle \hat{C}^{Ez}_\ell \right\rangle &= -\frac{4\pi}{\sqrt{\ell(\ell+1)}} \frac{(\ell-2)!}{(\ell+2)!} \int_{-\infty}^{+\infty} df \frac{1}{2} F^E_\ell (2\pi fr) S_h(f) \\
        &= -\frac{4\pi}{\sqrt{\ell(\ell+1)}} \frac{(\ell-2)!}{(\ell+2)!} \int_{0}^{+\infty} df \frac{F^E_\ell (-2\pi fr)+F^E_\ell (2\pi fr)}{2} S_h(f) \\
        &= -\frac{4\pi}{\sqrt{\ell(\ell+1)}} \frac{(\ell-2)!}{(\ell+2)!} \int_{0}^{+\infty} df\ \mathrm{Re}\left[F^E_\ell (2\pi fr)\right] S_h(f),
    \end{aligned}
\end{equation}
using the fact that~$\left[F^E_\ell(\tilde{x})\right]^* = F^E_\ell(-\tilde{x})$, finding in this way a real angular power spectrum.
An alternative fashion to reach the same conclusion is based on the observation that, in the long-distance limit,
\begin{equation}
    z_{\ell m }^{f} = - \frac{\sqrt{\ell(\ell+1)}}{2} E_{\ell m}^{f,r\to\infty}.
\end{equation}
Therefore, we have that
\begin{equation}
    \begin{aligned}
        C_\ell^{zE}(r) = \langle z_{\ell m}{E_{\ell m}^{f,r}}^* \rangle = - \frac{\sqrt{\ell(\ell+1)}}{2} \langle {E_{\ell m}^{f,r'\rightarrow\infty}}{E_{\ell m}^{f,r}}^* \rangle = - \frac{\sqrt{\ell(\ell+1)}}{2} C_\ell^{EE}(r,r'\rightarrow\infty).
    \end{aligned}
\end{equation}
Since~$C_\ell^{EE}(r,r')$ are real angular power spectra, as already argued above, it descends that the same holds for $C_\ell^{zE}(r)$.


\section{Properties of star and asteroid populations}
\label{app:population_properties}

Here we discuss more technical aspects of the implementation of our target observables.
Before proceeding forward, we want to note that none of the fitting formulas implemented have any background motivation, since our goal is just to obtain an observation-based number distribution.

\begin{figure}[ht]
    \centerline{
    \includegraphics[width=\linewidth]{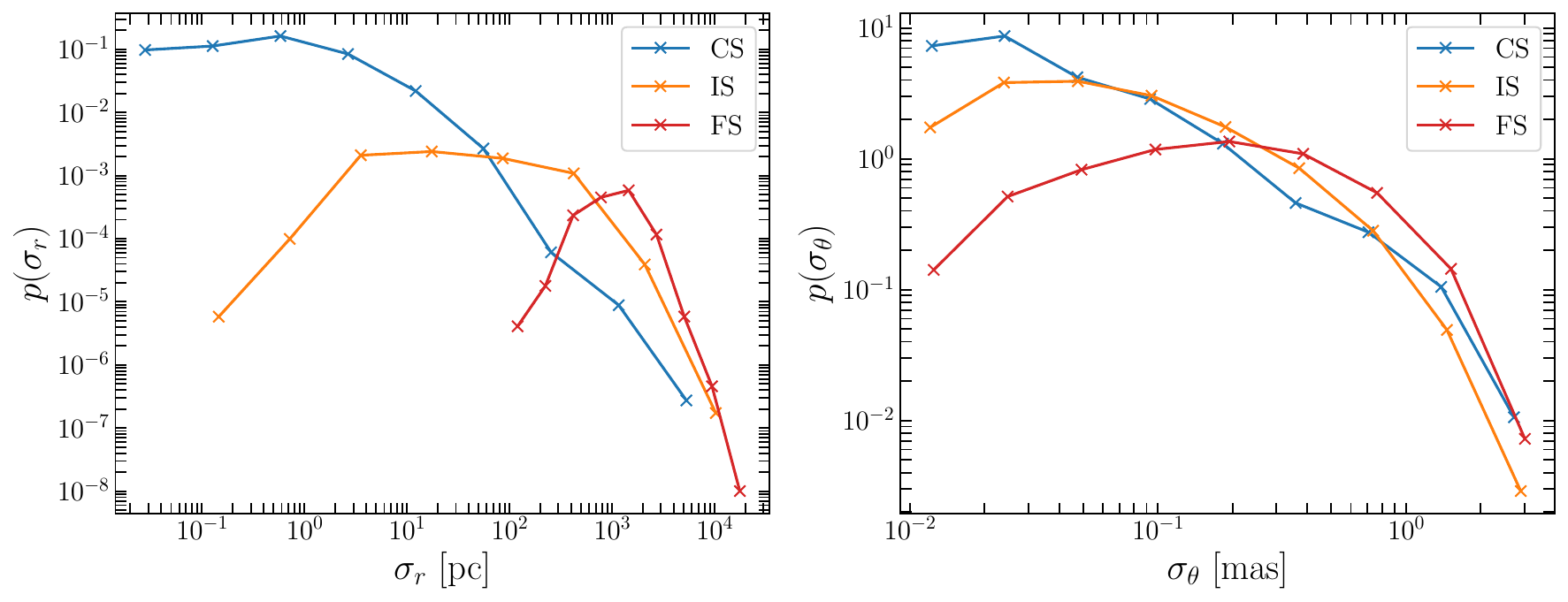}}
    \caption{Probability distribution functions for the distance (\textit{left panel}) and angular position (\textit{right panel}) errors for stars in the GAIA sample}
    \label{fig:stars_distance_angle_errors}
\end{figure}

For stars with Bayesian determination of the distance, we fit their radial distribution probability distribution function using
\begin{equation}
    \Phi_\mathrm{star}(r_\mathrm{med}) = \mathcal{N}^{-1}_\mathrm{star}
    \left\lbrace \begin{aligned}
        &\left( \frac{r_\mathrm{med}}{\rho_0} \right)^{\alpha_0} \qquad &r_\mathrm{min} \leq r_\mathrm{med} \leq \rho_1, \\
        &\left( \frac{\rho_1}{\rho_0} \right)^{\alpha_0} \left( \frac{r_\mathrm{med}}{\rho_1} \right)^{\alpha_1} \qquad &\rho_1 \leq r_\mathrm{med} \leq \rho_2, \\
        &\left( \frac{\rho_1}{\rho_0} \right)^{\alpha_0} \left( \frac{\rho_2}{\rho_1} \right)^{\alpha_1} e^{- \frac{\log^2 (r_\mathrm{med}/\rho_2)}{2\sigma^2} } \qquad &\rho_2 \leq r_\mathrm{med} \leq r_\mathrm{max}, \\
    \end{aligned} \right.
\label{eq:Phi_stars}
\end{equation}
where the values of the bestfit parameters are reported in table~\ref{tab:stars_fitting_parameters}.
By definition, the value of the normalization coefficient~$\mathcal{N}_\mathrm{star}$ is fixed by imposing the normalization condition.
Additionally, in figure~\ref{fig:stars_distance_angle_errors} we report the probability distribution functions for both the distance and angular position errors.

\begin{table}[ht]
    \centerline{
    \begin{tabular}{|c|c|c|c|c|c|c|}
        \hline
        $\rho_0\ [\mathrm{pc}]$ & $\rho_1\ [\mathrm{pc}]$ & $\rho_2\ [\mathrm{pc}]$ & $\rho_3\ [\mathrm{pc}]$ & $\alpha_0$ & $\alpha_1$ & $\sigma$ \\
        \hline
        \hline
        14 & 300 & 3700 & 32000 & 1.85 & 0.5 & 0.47 \\
        \hline
    \end{tabular}}
    \caption{Best-fit coefficients of the star radial number probability distribution function in equation~\eqref{eq:Phi_stars}.}
    \label{tab:stars_fitting_parameters}
\end{table}

On the other hand, regarding the three families of asteroids, we characterize their radial number probability distribution function using the fitting formulas
\begin{equation}
    \Phi_\mathrm{ast}(a) = \mathcal{N}^{-1}_\mathrm{ast} \left[ \frac{A_1}{\sqrt{2\pi \sigma^2_1}} e^{-\frac{1}{2}\left(\frac{a-\bar{a}_1}{\sigma_1} \right)^2} + \frac{A_2}{\sqrt{2\pi \sigma^2_2}} e^{-\frac{1}{2}\left(\frac{a-\bar{a}_2}{\sigma_2} \right)^2} + \frac{A_3}{\sqrt{2\pi \sigma^2_3}} e^{-\frac{1}{2}\left(\frac{a-\bar{a}_3}{\sigma_3} \right)^2} \right],
\label{eq:Phi_asteroids}
\end{equation}
where the fitting parameters are reported in table~\ref{tab:asteroids_fitting_parameters}.
As before, the value of the normalization coefficient~$\mathcal{N}_\mathrm{ast}$ is fixed by imposing the normalization condition.
The necessity of this structure is demanded by the presence of substructures in the three families, however, not all these parameters are, in fact, necessary for all the families of asteroids.

\begin{table}[ht]
    \centerline{
    \begin{tabular}{|c|c|c|c|c|c|c|c|c|c|}
        \hline
        Family & $A_1$ & $\bar{a}_1$ [AU] & $\sigma_1$ [AU] & $A_2$ & $\bar{a}_2$ [AU] & $\sigma_2$ [AU] & $A_3$ & $\bar{a}_3$ [AU] & $\sigma_3$ [AU] \\
        \hline
        \hline
        IMB & $1.06$ & $2.34$ & $0.09$ & $\varnothing$ & $\varnothing$ & $\varnothing$ & $\varnothing$ & $\varnothing$ & $\varnothing$ \\
        MMB & $0.47$ & $2.75$ & $0.06$ & $0.61$ & $2.60$ & $0.06$ & $\varnothing$ & $\varnothing$ & $\varnothing$ \\
        OMB & $1.38$ & $3.15$ & $0.13$ & $0.06$ & $2.87$ & $0.02$ & $-0.28$ & $3.25$ & $0.04$ \\
        \hline
    \end{tabular}}
    \caption{Best-fit coefficients of the star radial number probability distribution function in equation~\eqref{eq:Phi_asteroids}.
    The~$\varnothing$ symbol indicates that those sets of parameters are not needed to describe the asteroid radial probability.
    More explicitly, to describe the IMB, MMB, and OMB families we need one, two, and three Gaussian functions, respectively.}
\label{tab:asteroids_fitting_parameters}
\end{table}

\begin{figure}[ht]
    \centerline{
    \includegraphics[width=\linewidth]{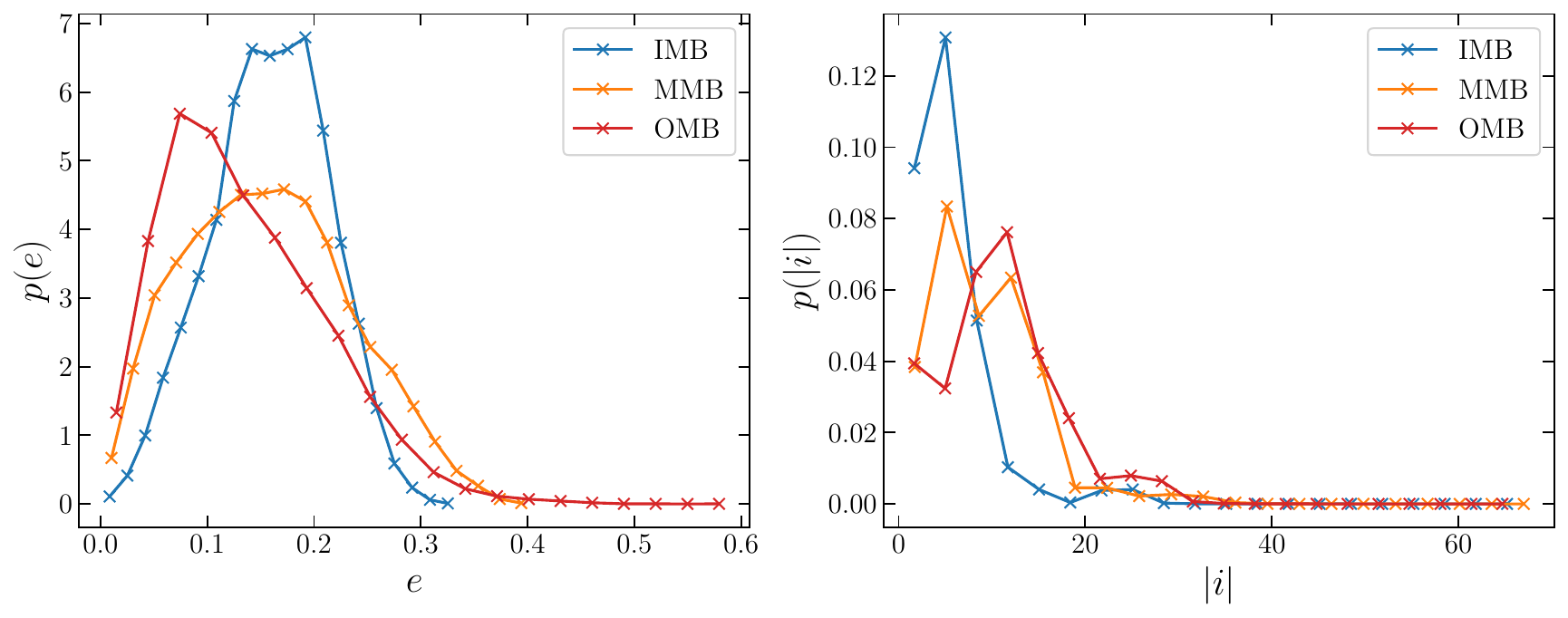}}
    \caption{Probability distribution function for eccentricity~$e$ (\textit{left panel}) and inclination~$i$ (\textit{right panel}) parameters for IMB, MMB, and OMB asteroids in the NASA Small-Body Database.
    Markers indicate the real catalog-based value.}
\label{fig:asteroid_eccentricity_inclination}
\end{figure}

Finally, we report in figure~\ref{fig:asteroid_eccentricity_inclination} the probability distribution functions of both the eccentricity and inclination of the three asteroids families.

\bibliography{refs}
\bibliographystyle{JHEP}

\end{document}